\documentclass[lettersize,journal]{IEEEtran}
\usepackage{amsmath}
\usepackage{amssymb}
\usepackage{booktabs} % For formal tables
\usepackage{multirow}
\usepackage{float} % strict floating figures
\usepackage[linesnumbered,ruled]{algorithm2e} % For algorithms
\usepackage[svgnames]{xcolor}

\usepackage{wrapfig}
\usepackage{rotating}
\usepackage{subcaption}
\usepackage[percent]{overpic} % for overlaying text on images
\usepackage{tikz}
\usetikzlibrary{arrows.meta} % for control over arrows
\usetikzlibrary{math} % for arithmetic computations
\usetikzlibrary{decorations.pathreplacing,decorations.pathmorphing} % for interface diagram
\usepackage{hyperref}

\SetAlFnt{\normalsize}
\SetAlCapFnt{\normalsize}
\SetAlCapNameFnt{\normalsize}
\SetAlCapHSkip{0pt}
\IncMargin{0pt}

\DeclareMathAlphabet{\mathsfit}{T1}{\sfdefault}{\mddefault}{\itdefault}
\SetMathAlphabet{\mathsfit}{normal}{T1}{\sfdefault}{\mddefault}{\itdefault}
\DeclareMathAlphabet{\mathsfbdit}{T1}{\sfdefault}{\bfdefault}{\sldefault}
\SetMathAlphabet{\mathsfbdit}{bold}{T1}{\sfdefault}{\bfdefault}{\sldefault}

\newcommand{\Description}[2][]{} % Silence errors from ACM latex docs.
\newcommand{\shortcite}[1]{\cite{#1}}

\newcommand{\new}[1]{{#1}}
\def\newTitle{}

\newcommand{\newTvcg}[1]{{#1}}
\def\newTvcgTitle{}

\def\genv#1{\boldsymbol{\mathrm{#1}}} % generalized coordinate space vector
\def\tt{t} % time variable
 
\def\uu{\genv{u}} 
\def\cc{\genv{c}} 
\def\q{\genv{q}} 
\def\vv{\genv{v}} 
\def\M{\genv{M}} 
\def\K{\genv{K}} 
\def\D{\genv{D}} 
\def\T{\genv{T}}
\def\B{\genv{B}}
\def\J{\genv{J}}
\def\f{\genv{f}} 
\def\dd{\genv{d}}

\def\y{\genv{y}}
\def\rr{\genv{r}}

\def\p{\genv{p}}

\def\blambda{\genv{\lambda}}
\def\eeta{\genv{\eta}}
\def\mmu{\genv{\mu}}

\def\R{\mathbb{R}}

\def\BE{\scriptscriptstyle \mathrm{BE}}
\def\TR{\scriptscriptstyle \mathrm{TR}}
\def\BDF2{\scriptscriptstyle \mathrm{BDF2}}
\def\SDIRK2{\scriptscriptstyle \mathrm{SDIRK2}}

\def\Jac{\mathbf{J}}
\DeclareMathOperator*{\argmax}{argmax}

\def\myeqref#1{Eq.~\eqref{#1}}
\def\figref#1{Figure~\ref{#1}}
\def\algref#1{Algorithm~\ref{#1}}

\def\tabref#1{Table~\ref{#1}}

\def\secref#1{Section~\ref{#1}}

\title{Implicit frictional dynamics with soft constraints \newTitle \newTvcgTitle}

\author{%
  Egor Larionov,
  Andreas Longva,
  Uri M. Ascher,
  Jan Bender, and
  Dinesh K. Pai
  \thanks{Manuscript created December 2023; This work was supported by the Natural Sciences and Engineering Research Council Discovery Grant, 
  and the Deutsche Forschungsgemeinschaft (DFG, German Research Foundation) -- Project number BE 5132/5-1.}
}

\begin{document}

\maketitle

\begin{abstract}
  Dynamics simulation with frictional contacts is important for a wide range of
  applications, from cloth simulation to object manipulation. Recent methods
  using smoothed \new{lagged} friction forces have enabled robust and differentiable
  simulation of elastodynamics with friction. However, the resulting frictional
  behavior can be inaccurate and may not converge to analytic
  solutions.
  \new{Here we evaluate the accuracy of lagged friction models
  in comparison with implicit frictional contact systems.
  We show that major inaccuracies near
  the stick-slip threshold in such systems are caused by lagging of friction
  forces rather than by smoothing the Coulomb friction curve.}
  \new{Furthermore, we demonstrate how systems involving implicit or lagged
  friction can be correctly used with higher-order time integration and
  highlight limitations in earlier attempts.
  We \newTvcg{demonstrate how} to exploit forward-mode automatic differentiation to
  simplify and, in some cases, improve the performance of the inexact Newton
  method.
  Finally, we show that other complex phenomena can also be simulated
  effectively while maintaining smoothness of the entire system.  We extend our
  method to exhibit stick-slip frictional behavior and preserve volume on
  compressible and nearly-incompressible media using soft constraints.}
\end{abstract}

% \keywords{dry friction, contact, elasticity}

% \graphicspath{{.}} % where to search for the images

\begin{figure*}[t]
  \centering
    \includegraphics[width=\linewidth,trim={0 0 0 0},clip]{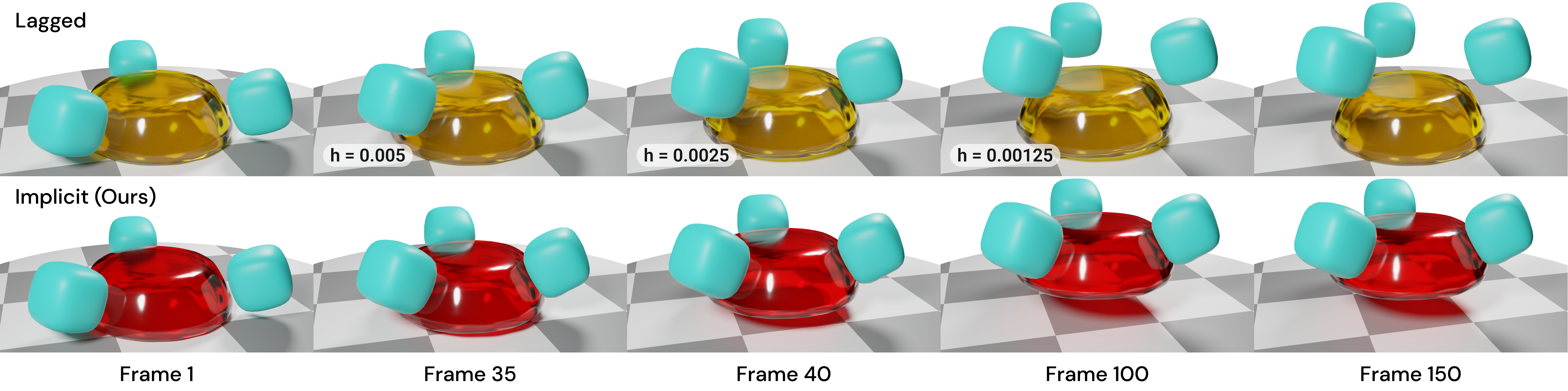}%
    \Description[Three soft pads picking up an upside-down bowl]{Renders of three soft pads lifting an upside-down bowl}
      \caption{An upside-down bowl is lifted using 3 soft pads via friction.
      The bowl is simulated using the lagged friction model from \shortcite{li20}
      (top row) and our fully implicit method
      (\myeqref{eq:be_residual}, bottom row).
      The chosen frames between 1 and 150, are frames
      35, 40 and 100, at which point the bowl slips out for time steps $h =
      0.005$ s, $0.0025$ s, and $0.00125$ s respectively for the lagged method.
      Using fully implicit integration,
      the bowl sticks even at the largest time step $h = 0.005$ s as shown.
    For the pads, $\rho = 1000$ kg/m${}^3$ (density),
    $E = 600$ KPa (Young's modulus), and $\nu = 0.49$ (Poisson ratio).
    For the bowl, $\rho = 400$ kg/m${}^3$, $E = 11000$ KPa and $\nu = 0.1$.
    The friction coefficient is set to $\mu = 0.65$ between the two.
    See \figref{fig:bowl_grasp_height} for the plot of the bowl height in all tested configurations.
    \label{fig:bowl_grasp}}
\end{figure*}

\section{Introduction} \label{sec:intro}

% General Intro to friction and contact
Modern simulation pipelines in computer graphics and engineering involve contact handling.
This ensures that simulated objects do not interpenetrate each other as they interact.
During this interaction, friction ensures that solid objects are held in place or are otherwise
limited in how they move. This work focuses on the accuracy and effectiveness of smooth friction models.

% Nonlinear equations
In recent years we have seen a resurgence of promising work in developing more
robust methods for simulating frictional contact in computer graphics.
This problem is particularly difficult since friction and contact cannot be simultaneously
described by a single energy potential \cite{desaxce98}.  This precludes formulating the frictional contact
problem as a single energy minimization. However, energy minimization used for solving
discretized ordinary differential equations (ODEs) over each time step has remained popular in graphics,
due to its flexibility and robustness characteristics. Unfortunately, optimization based solvers
used for dynamics equations require specialized algorithms for handling frictional contacts,
which necessarily produces drawbacks in accuracy or robustness.

In this work, we demonstrate failure cases in popular optimization based frictional contact solvers
and propose an alternative method for solving elastodynamic problems
with frictional contacts that is simple to implement and accurate in comparison.
By resolving the sliding frame and contact forces implicitly when computing friction, our method
can produce more accurate friction behavior, and it requires
no additional iterations or specialized 
mechanisms for coupling friction, contact and elasticity.

% Smoothness
Frictional contact is traditionally modeled as a non-smooth problem requiring sophisticated
tools. In particular, non-smooth integrators, root finding or optimization techniques are needed
for handling inclusion terms in the mathematical model. This
drastically complicates the problem and substantially limits the number of solution approaches.
While non-smoothness is required to guarantee absolute sticking, it is not generally
necessary if simulations are limited in time.
In fact, when observed on a microscale, even dry friction responds continuously
to changes in velocity \cite{wojewoda08}.
Accordingly, we adopt a smooth friction formulation. We show that when applied
fully implicitly in the equations of motion, it can produce predictable
sticking. In contrast, our study shows that the popular approach of lagging
friction causes inaccurate and time step dependent sticking behavior.
In particular, we demonstrate
the importance of evaluating the sliding basis (defined in Section \ref{sec:friction}) and contact forces implicitly 
for accurate friction simulation.

Finally, to maintain smoothness of the entire problem we employ a
penalty force for contact resolution and
a smooth implicit surface model proposed by Larionov et al. \shortcite{larionov21} for representing the contact surface.
We also show how additional soft constraints can be added to the system for controlling the
volume of an object.

Maintaining smoothness of the entire system makes it easier to solve and allows derivatives to be propagated
through each step of the simulation. However, full differentiable simulation is outside the scope of this work.

% Contributions
In summary, our primary contributions are
\begin{itemize}
	\item \new{A comparison between lagged and fully implicit friction models focusing on frictional accuracy.}
	\item \new{A stable implementation of high-order time integration for
	frictional contact problems that is applicable to both lagged and fully
	implicit formulations.}
	\item \new{A novel smooth Stribeck friction model used for modeling stick-slip
	effects due to the microstructure of contacting surfaces.}
	\item A physically-based volume change penalty for controlling compressibility in compressible and nearly incompressible regions.
	\item A \new{simple} adaptive penalty stiffening strategy for effectively resolving interpenetrations with penalty-based contact methods.
\end{itemize}

Furthermore, to better characterize the instability of single point frictional contacts, we
present an eigen-analysis of a 2D point contact subject to friction in
Section I of the supplemental document.

\section{Related Work} \label{sec:related_work}

Simulating the dynamics of deformable elastic objects \cite{terzopoulos87} and cloth
\cite{baraff98} has been an active area of research in graphics for decades.
Methods requiring accuracy often reach for a finite element method (FEM) \cite{sifakis12}, while methods
aiming for performance reach for position-based techniques \cite{muller07, macklin16, bender17},
or projective-dynamics \cite{bouaziz14}. Our work targets simulation accuracy,
and thus stays close to the tried and tested FEM.

\paragraph{Frictional Contact}
In recent years, lots of attention has been dedicated towards robust and
accurate contact and friction solutions. Earlier works in graphics developed
non-smooth methods to resolve contact and friction forces at the end of the
time step obtaining solutions faithful to the Coulomb model.
Kaufman et al. \shortcite{kaufman08} proposed a predictor-corrector method to solve 
for friction and contact separately from elasticity equations. This was later extended to high-order time integration \cite{chen17} and
separately to full FEM simulations \cite{larionov21} while maintaining the decoupling. Other methods reformulate
the problem as a non-smooth root finding problem \cite{bertails11, daviet11, li18, macklin19} or using proximal algorithms
\cite{erleben17}.  More recently, more attention was brought towards modeling friction as a smoothly changing force
at the stick-slip limit \cite{geilinger20,li20}.  This allows each simulation step to remain differentiable.
Geilinger et al. \shortcite{geilinger20} favored a more traditional root-finding solver combining friction and contact forces with 
elastic equations of motion. In contrast, Li et al. \shortcite{li20} proposed a robust optimization
framework to solve for contact and lagged friction forces.
Unfortunately, even with multiple iterations, the lagged friction approach might not converge to an accurate
friction solution (cf. \figref{fig:bowl_grasp}), which is especially noticeable in sticking configurations close to the slip threshold.
Furthermore, their proposed method for higher-order integration is not applied to the contact solve causing
instabilities.
In this work, we demonstrate and address these shortcomings using a solution method that favors friction
accuracy at the cost of some robustness, while maintaining smoothness of the system.
\new{While non-smooth methods for frictional contact are interesting and
have been physically validated \cite{chen17}, our work is focused solely on smooth methods.}

\paragraph{Higher-order integrators for contact problems}
Most contact formulations, especially those formulated in terms of constraints, intrinsically rely on a particular choice of time discretization,
which is usually backward Euler (BE).
However, the highly dissipative characteristics of BE have motivated the use of higher-order schemes like TR-BDF2 or SDIRK2, which preserve high energy dynamics while maintaining stability~\cite{loschner20, ascher21}.
A benefit of smooth contact models based on penalty or barrier functions is that both normal and friction forces are defined with explicit formulas, as opposed to implicitly defined through constraints. % and Lagrange multipliers. (Egor) omitting for brevity
This makes it possible to apply higher-order integrators directly, as demonstrated by Geilinger et al. \shortcite{geilinger20} for BDF2.
Li et al. \shortcite{li20} applied trapezoid rule (TR), however it is only applied to non-contact forces, which \new{we show} causes stability issues.
Brown et al. \shortcite{brown18} focus on first-order methods and apply TR-BDF2
to a non-smooth optimization-based contact model with lagged friction, though
this has unclear implications for high-accuracy second-order methods.
\new{Here, we show how high-order methods can be applied to systems with lagged or implicit friction methods.}

\paragraph{Volume preservation}
Many solids exhibit volume preserving behavior. We focus primarily on inflated
objects like tires and sports equipment (e.g. sports balls), as well as nearly
incompressible objects like the human body.
Inflated objects are typically simulated using soft constraints
\newTvcg{\cite{bonet00, skouras12, skouras14, bender17, panetta21}},
where \newTvcg{in effect the volume change of an object is penalized assuming Boyle's law for ideal gasses}.
These methods are effective, however, their physical accuracy is rarely questioned.
Incompressible or nearly-incompressible materials are often modeled with stiff
Poisson's ratios \cite{smith17} or hard volume preservation constraints \cite{sheen21}.
In contrast to previous work, we propose a unified physically-based penalty formulation for
volume preservation that models both compressible and nearly-incompressible objects
using a single penalty controlled by a physical compression coefficient. \newTvcg{Additionally, we
illustrate how the volume preservation models differ, indicating the expected trade-offs in each model}.

\section{Method}
\label{sec:formulation}

Here we develop the equations involved in solving for the
motion of deformable objects subject to frictional contacts.

A set of generalized coordinates $\q(\tt) \in \R^m$
(e.g. stacked vertex positions of a FEM mesh in 3D),
represents a system of solids moving through
time $\tt$. A sparse symmetric positive definite (SPD) $m \times m$ matrix $\M$ denotes generalized mass.
In the following we omit the time dependence for brevity, but later reintroduce it when discussing time
discretization schemes.

Using dot notation for time derivatives and with generalized velocities
$\vv = \dot{\q}$ we can write the force balance equation as
\begin{subequations}%
\begin{align}%
  \M\dot{\vv} & = \f(\q, \vv)  \label{eq:eom}                                               \\
  \f(\q, \vv) & = \f_e(\q) + \f_d(\q,\vv) + \f_c(\q) + \f_f(\q, \vv) + \f_g(\q) \label{eq:f},
\end{align}%
\end{subequations}%
where $\f_e$ are elastic forces, $\f_d$ are damping forces, $\f_c$ are contact forces, $\f_f$ is friction,
and $\f_g$ are external forces such as gravity.

\subsection{Elasticity and Damping}
Elastic forces are typically derived from a configuration dependent energy potential $W(\q)$ as
$\f_e(\q) = -\frac{\partial}{\partial \q}W(\q)$.
The elastic potential $W$ can be defined by the classic linear, neo-Hookean,
StVK or Mooney-Rivlin models \cite{ciarlet83}, or even a data-driven model \cite{wang11}.
We will focus on neo-Hookean materials for both solids and cloth.
We then define the \emph{stiffness matrix}
$\K(\q) = -\frac{\partial}{\partial \q}\f_e(\q)$, which dictates how resistant an
object is to deformation. For nonlinear models like neo-Hookean elasticity,
$\K$ may be indefinite, which is important to know when picking an appropriate linear solver.

Damping forces are often defined by
$\f_d(\q, \vv) = -\D(\q)\vv$,
where $\D$ is a symmetric, and usually positive semi-definite, matrix.
We use the Rayleigh damping model for simplicity
where $\D = \alpha \M + \beta \K$ for some constants $\alpha, \beta \geq 0$.

\subsection{Contact}
Traditionally contact constraints have been formulated with a
nonnegativity constraint on some ``gap'' function $\dd(\q)$ that roughly
determines how far objects are away from each other.  This function $\dd$ may be
closely related to a component-wise signed distance function, however,
generally it merely needs to be continuous, monotonically increasing in the
direction of separation, and constant at the surface. We define $d_i$ for each
potential contact point $i$ such that $\dd = (d_1, d_2, \dots, d_n)$, where $n$
is the total number of potential contacts.
Here we use the contact model of Larionov et al. \shortcite{larionov21},
where surface vertices of one object are constrained to have non-negative
potential values when evaluated against a smooth implicit function $\dd$ closely approximating
the surface of another object.
Interestingly, if we allow objects some small separation tolerance at
equilibrium, we can reformulate this constraint as an \emph{equality}
constraint by using a soft-max \cite{geilinger20} or a truncated log barrier
\cite{li20} function. These types of equality constraints greatly simplify the
contact problem and have shown tremendous success in practice.

One disadvantage of log-barrier formulations is that the initial state must be
free of collisions prior to the optimization step in order to avoid infinite energies and
undefined derivatives. In the absence of thin features or risk of tunneling, it is
sufficient to use a simple penalty function to resolve interpenetrating geometry.
In this work we choose to use penalty-based contacts for simplicity, however,
our formulation is fully compatible with a log-barrier method coupled with continuous collision detection (CCD) as proposed
by Li et al. \shortcite{li20}.
The idea is to help our solver guide interpenetrating meshes out of intersecting
configurations. We define a cubic contact penalty by
\begin{align*}
  \newTvcg{b(x; \delta, \kappa) = \kappa \left\{\begin{array}{ll}
    -\frac 1 \delta(x-\delta)^3 & \text{ if } x < \delta \\
    0                            & \text{ otherwise,}
  \end{array}\right.}
\end{align*}
where $\delta > 0$ is the thickness tolerance and $\kappa > 0$ is a contact
stiffness parameter that will need to be automatically increased to ensure that
no surface vertices of one object are penetrating the implicit surface of another
at the end of the time step.  Here $b$
corresponds to the first non-zero term in the Taylor expansion of the truncated
log-barrier used by Li et al. \shortcite{li20}, but unlike the log-barrier it is well-defined
also for negative arguments.  The penalty is applied to each pair of contact
points with distance $d_i$ giving us an aggregate contact energy
  \begin{align*}
   W_c(\dd) = \sum_{i=1}^n b(d_i; \delta, \kappa).
\end{align*}
Now the contact force can be written simply as the negative energy derivative
\begin{align}
  \f_c(\q)^{\top} = -\frac{\partial W_c}{\partial \q} = \blambda(\q)^\top \frac{\partial \dd}{\partial \q}, &  & \text{where} &  &
  \blambda(\q)^\top = -\frac{\partial W_c}{\partial \dd}
  \label{eq:contact_force}
\end{align}
is the stacked vector of contact force magnitudes. In
effect our contact formulation enforces the equality constraint
$W_c(\dd(\q)) = 0$.

\subsection{Friction} \label{sec:friction}

We define the \emph{contact Jacobian} $\J_c(\q)$ and tangent basis $\B(\q)$
over all potential contact points as in \cite{larionov21}.
Then $\T(\q) = \J_c(\q)^{\top}\B(\q)$ is the $m\times n$ matrix defining the \emph{sliding basis}
\cite{li20}. In short, this matrix maps forces in contact space
to generalized forces in configuration space.

We can now derive the smoothed friction force \cite{li20, geilinger20} from first principles.
For each contact $i$, the maximum dissipation principle (MDP) postulates that friction force ought to
maximally oppose relative velocity
\begin{align}
  \f_{f,i}(\vv; \mu) = \argmax_{\|\y\| \leq \mu \lambda_i}( -\bar\vv_i^{\top} \y ), \label{eq:mdp}
\end{align}
where $\mu$ is the coefficient of friction, which limits the friction
force\footnote{Unless otherwise specified, $\|\cdot\|$ refers to the Euclidean
norm.} and $\bar\vv_i \in \R^2$ is the relative tangential velocity at contact
point $i$.  The contact force magnitude $\lambda_i$ is the $i$th element of
$\blambda$ as defined in~\myeqref{eq:contact_force}.  We can solve~\myeqref{eq:mdp}
explicitly with an inclusion
\begin{align}
  \f_{f,i}(\vv; \mu) \in -\mu \lambda_i\left\{\begin{array}{ll}
    \{\bar\vv_i/\|\bar\vv_i\| \}                      & \text{ if } \|\bar\vv_i\| > 0 \\
      \{\uu \in \R^2\ :\  \|\uu\| \leq 1 \} & \text{ otherwise}.
  \end{array}\right. \label{eq:coulomb}
\end{align}
This is commonly referred to as Coulomb friction. Unfortunately, the non-smoothness around
$\|\bar\vv_i\| = 0$ calls for non-smooth optimization or root-finding techniques
\cite{kaufman08, bertails11, erleben17}, making this problem numerically challenging.
Another disadvantage of non-smoothness is that it greatly complicates differentiation of the
solver, which can be critical for solving inverse problems efficiently. We opt to approximate Coulomb
friction using a smoothed model \cite{li20,geilinger20}. Since most animations
call for relatively short time frames, we typically do not require absolute
sticking. Interestingly, smooth friction models have been proposed in older
engineering literature \cite{kikuuwe05, wojewoda08, awrejcewicz08} to improve
hysteretic behavior and alleviate numerical difficulties.
A simple smoothing \newTvcg{\cite{geilinger20}} of \myeqref{eq:coulomb} can be written as
\begin{align}
  \f_{f,i}(\vv; \mu) \approx -\mu\lambda_i s(\|\bar\vv_i\|) \eeta(\bar\vv_i),
\end{align}
where $\eeta : \R^2 \to \R^2$ defines the per-contact nonlinearity
\begin{align}
  \eeta(\bar\vv_i) = \left\{\begin{array}{ll}
    \bar\vv_i/\|\bar\vv_i\| & \text{ if } \|\bar\vv_i\| > 0 \\
    0                     & \text{ otherwise, }
  \end{array}\right. \label{eq:nonlinearity}
\end{align}
and the function $s$ defines the pre-sliding transition. A popular $C^1$ option
for $s$ as depicted in \figref{fig:smoother} for different values of $\epsilon$,
is
\begin{align}
  s(\newTvcg{v}; \epsilon) = \left\{\begin{array}{ll}
    \frac{2\newTvcg{v}}{\epsilon} - \frac{\newTvcg{v}^2}{\epsilon^2} & \text{ if }\newTvcg{v}<\epsilon \\
    1                                            & \text{ otherwise. }
  \end{array}\right.\label{eq:smoother}
\end{align}
\new{Simpler and smoother functions exist, however this choice is convenient since
$\epsilon$ directly controls the sliding velocity tolerance during sticking.

We can then easily express more complex friction models by substituting
$\mu \lambda_i s(\|\bar\vv_i\|)$ with a velocity dependent function $c(\newTvcg{v}, \lambda_i; \mmu)$ of multiple coefficients $\mmu$.
For instance, Stribeck and viscous friction \cite{armstronghelouvry93} can be
expressed with $\mmu = (\mu_d, \mu_s, \mu_v)$ representing
dynamic, static and viscous friction coefficients respectively:
\begin{align}
  \f_{f,i}(\vv; \mmu) &= - c(\|\bar\vv_i\|, \lambda_i; \mmu) \eeta(\bar\vv_i), \notag \\
  c(\newTvcg{v}, \lambda_i; \mmu) &= (\mu_d + (\mu_s - \mu_d) g(\newTvcg{v}/v_s))s(\newTvcg{v}; \epsilon) \lambda_i + \mu_v \newTvcg{v} \label{eq:stribeck}\\
  g(x) &= \left\{\begin{array}{ll}
    (2x + 1)(x-1)^2 & \text{ if } x < 1 \notag \\
    0 & \text{ otherwise, } \notag
  \end{array}\right.
\end{align}
where $g$ is a compact Gaussian approximation, and $v_s$ is Stribeck velocity, which
defines how gradually friction decays from static into dynamic as velocity
increases. \figref{fig:stribeck} illustrates how the different parameters in
\myeqref{eq:stribeck} control the overall curve. In general, $v_s$ should be
larger than $\epsilon$, where values close to $\epsilon$ may lower the observed/effective static friction force.
Setting $v_s = 10\epsilon$ works well for modeling common dry friction when $\epsilon$ is small.
The Stribeck component is a useful tool for introducing static friction into the
smoothed friction model. Our example in \secref{sec:tire_wrinkle} shows how
this model can introduce stick-slip behavior from the real world.}

\new{
\begin{figure}
  \centering
  \begin{subfigure}{0.48\linewidth}
    \includegraphics[width=\linewidth, trim={10 10 10 10}]{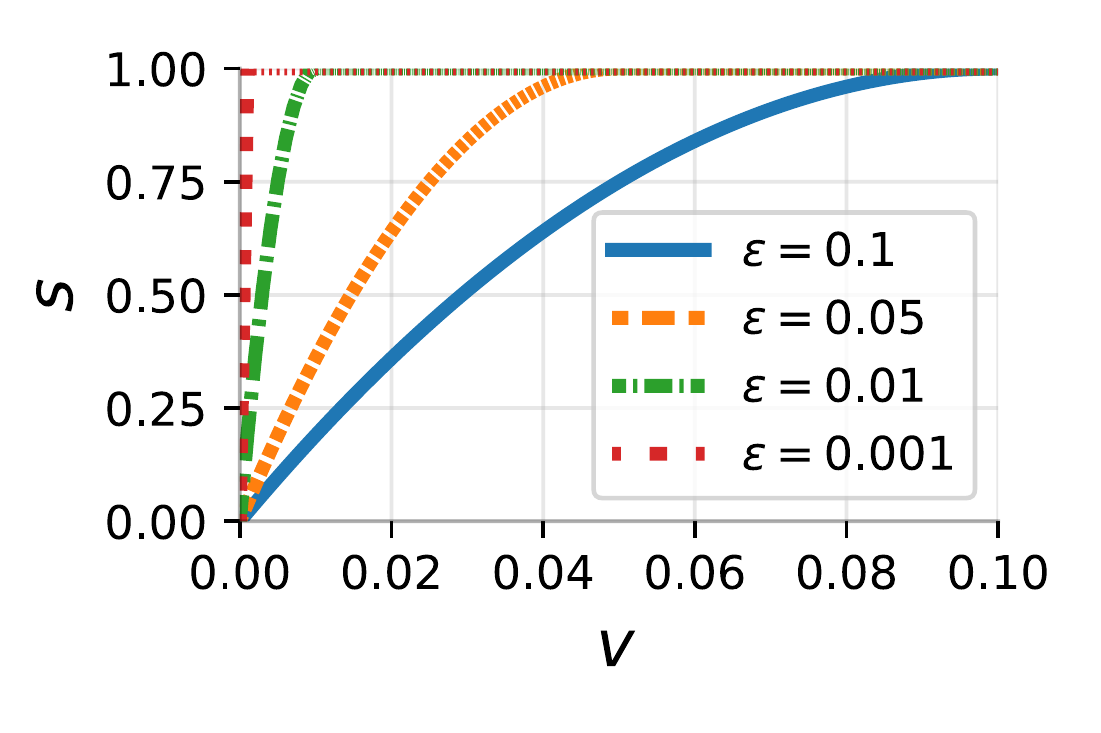}
    \caption{Pre-sliding transition $s$ as defined in \eqref{eq:smoother}. The function is plotted for different values of $\epsilon$, which control the velocity error tolerance. \label{fig:smoother}}
  \end{subfigure}\hfill
  \begin{subfigure}{0.48\linewidth}
    \includegraphics[width=\linewidth, trim={10 10 10 10}]{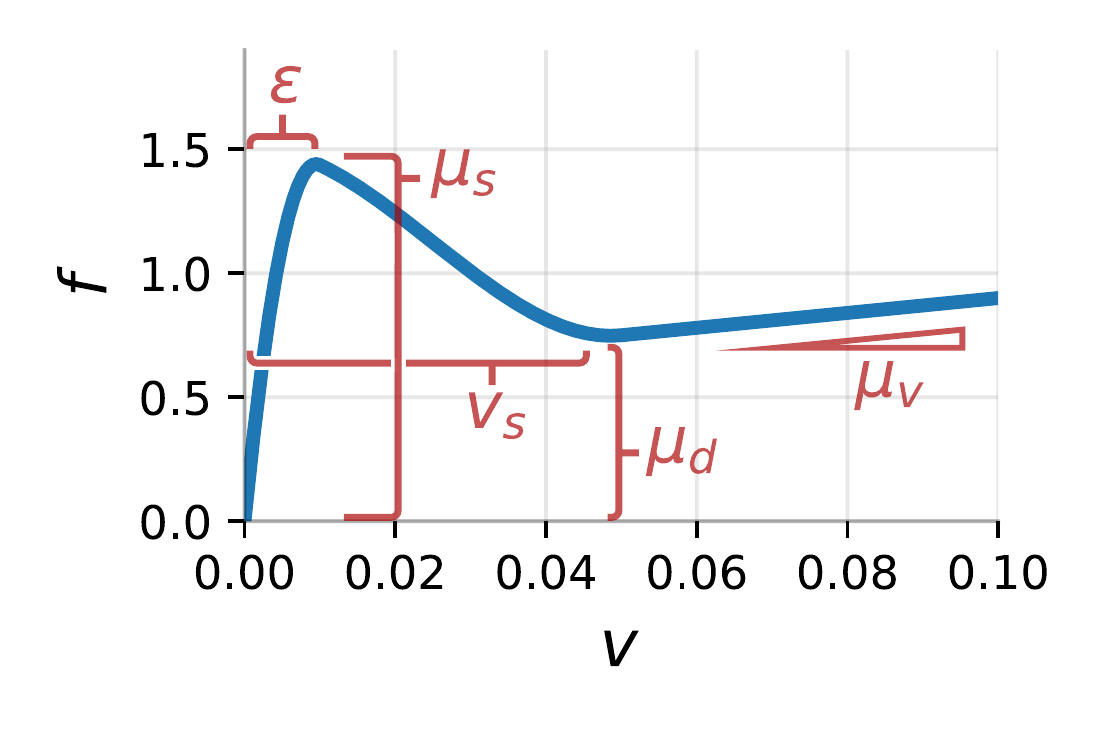}
    \caption{Stribeck and viscous friction curve $f$ as defined in
    \eqref{eq:stribeck}. The effect of each of the material properties in $\mmu$
    and parameters $\epsilon$ and $v_s$ are labelled.\label{fig:stribeck}}
  \end{subfigure}
  \caption{Components of the friction model.}\label{fig:smoothers}
\end{figure}
}

We can express the nonlinearity in \myeqref{eq:nonlinearity} as a function over all (stacked) relative velocities
$\bar\vv \in \R^{2n}$ using a diagonal block matrix
\begin{align*}
  \mathbf{H}(\bar{\mathbf{v}}) = \begin{bmatrix}
    \eeta(\bar{\mathbf{v}}_1) & & \\
    & \ddots & \\
    & & \eeta(\bar{\mathbf{v}}_n)
  \end{bmatrix}.
\end{align*}
Then the total friction force can be written compactly as
\begin{align}
  \f_f(\q,\vv) = -\T(\q)\mathbf{H}(\T(\q)^{\top}\vv)\cc(\q, \vv), \label{eq:friction_force}
\end{align}
\new{where $\cc_i = c(\|\bar\vv_i\|, \lambda_i; \mmu)$ is the $i$th component of the stacked vector
of friction force magnitudes.}

\subsection{Volume change penalty}

In soft tissue simulation, resistance to volume change is typically controlled
by Poisson's ratio.  This, however, assumes that the simulated body is
homogeneous and void of internal structure.  For more complex structures like
the human body, a zonal constraint is a more suitable method to enforce
incompressibility \cite{sheen21}.  Compressible objects, however, require a
different method altogether.  In this section we propose a physically-based and
stable model to represent compressible and nearly incompressible objects.  In
particular, we want to efficiently model inflatable objects like balloons, tires
and sports balls, as well as nearly incompressible objects like the human body
or other organic matter.

We start from the isothermal compression coefficient \cite[Section~5.3]{mandl71} defined by
\begin{align}
  \kappa_{v} = -\frac{1}{V}\left(\frac{\partial V}{\partial P}\right)_T, \label{eq:compressibility}
\end{align}
where $V$ is the volume of interest\footnote{For instance a region occupied by FEM elements or the volume of a watertight triangle mesh.}, $P$ is internal pressure and the $T$ subscript indicates
that temperature is held constant.
For compressible continua like air in normal conditions, which behaves like an ideal gas,
$\kappa_{v} = 1/P$. For nearly incompressible continua like water at room temperature,
$\kappa_{v} \approx 4.6\times 10^{-5}$ atm${}^{-1}$ is relatively constant.
% Compressibility of water sources:
% http://hyperphysics.phy-astr.gsu.edu/hbase/Tables/compress.html
% https://metalurji.mu.edu.tr/Icerik/metalurji.mu.edu.tr/Sayfa/MME%202009%20-%20Course3%20PVTRelationshipofFluids.pdf
Assuming rest volume $V_0$ and initial pressure $P_0=1$ atm, we can derive the work $W$
needed to change the volume of the container to $V$. For an ideal gas $PV$ is constant, which gives
\begin{align}
  W_{\mathit{ig}}(V) =  P_0\left(V - V_0\left(1 + \ln\frac{V}{V_0}\right)\right).
  \label{eq:ig_exact_volume_penalty}
\end{align}
For a nearly incompressible continuum, $\kappa_v$ is constant, which yields
\begin{align}
  W_{\mathit{ni}}(V) = \frac{1}{\kappa_v} \left(V_0 - V \left(1 - \ln\frac{V}{V_0}\right)\right).
  \label{eq:nif_exact_volume_penalty}
\end{align}
For details of the derivation see Section~II in the supplemental document.

\begin{figure}[h]
  \centering
  \includegraphics[width=\linewidth]{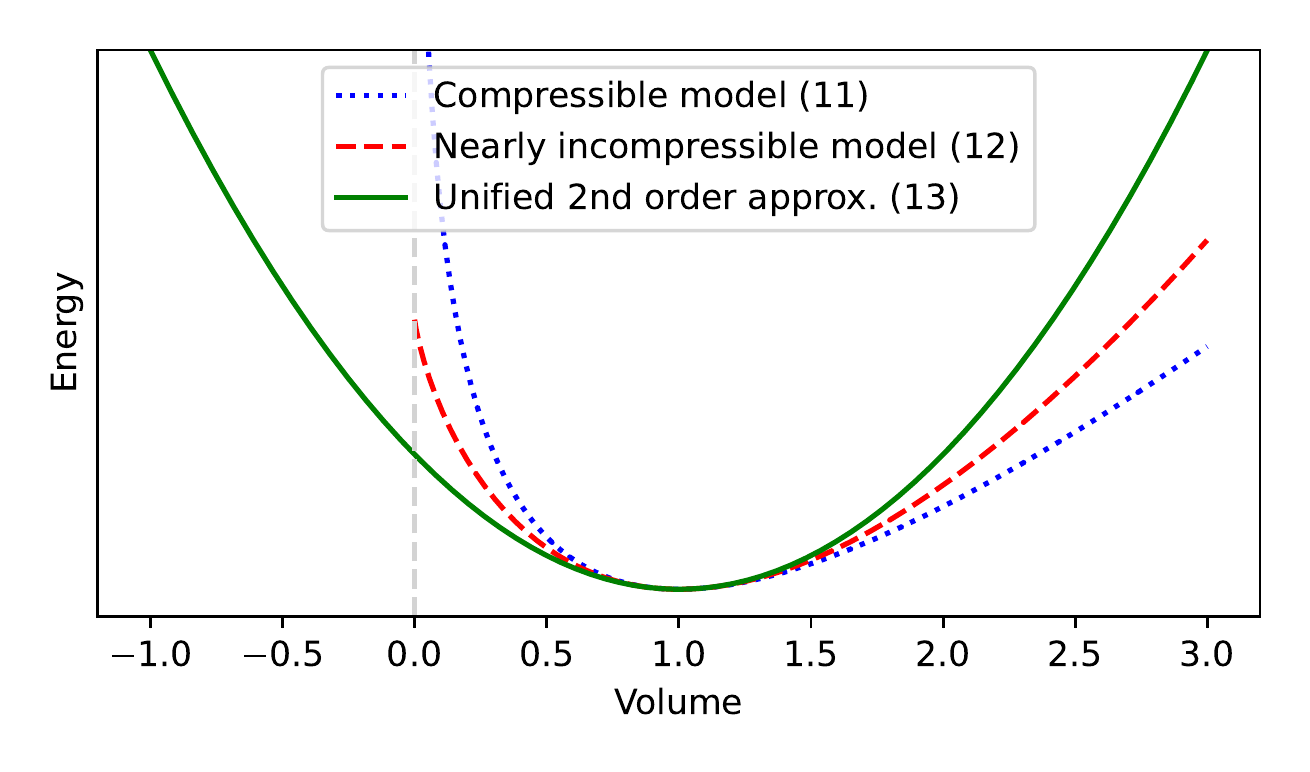}
  \caption{\emph{Volume change energy.} The energy (negative of work) is
  plotted for the compressible model in \myeqref{eq:ig_exact_volume_penalty} (dotted curve), 
  the nearly incompressible model in \myeqref{eq:nif_exact_volume_penalty} (dashed curve) and the 2nd
  order approximate model in \myeqref{eq:approx_volume_penalty} (solid curve). Here $V_0 = 1$ m${}^3$, $\kappa_v = 1$
  atm${}^{-1}$, and $P_0 = 1$ atm.
  The quadratic model approximates both cases, but is ultimately too weak for
  excessive compression yet too strong during large expansion.
  Depending on the use case, it may be necessary to model one of 
  Eqs. \eqref{eq:ig_exact_volume_penalty} or \eqref{eq:nif_exact_volume_penalty} directly.
  \label{fig:volume_change_penalty}}
  % Online version
  % https://www.desmos.com/calculator/encwlpbwwf
\end{figure}

Unfortunately, both models are undefined for negative volumes, which
can easily lead to configurations with undefined penalty forces.
Taking the second-order approximation of \myeqref{eq:nif_exact_volume_penalty} gives us
\begin{align}
  W_2(V) = \frac{(V - V_0)^2}{2V_0\kappa_v}, \label{eq:approx_volume_penalty}
\end{align}
which coincides with the second-order approximation of \myeqref{eq:ig_exact_volume_penalty} when $\kappa_v = 1$.
Thus, our second-order model approximates both compressible and nearly incompressible continua well for small changes in volume
as shown in \figref{fig:volume_change_penalty}.
For larger changes in volume, we recommend modeling
\myeqref{eq:ig_exact_volume_penalty} directly, since it also approximates \myeqref{eq:nif_exact_volume_penalty} well
and volume changes are not significant in nearly incompressible continua.

To alleviate the approximation error for scenarios that involve more compression
(such as in \figref{fig:ball_100mph}), we recommend decreasing $\kappa_v$
to produce stronger restorative forces.

The penalty force is then given directly by the negative derivative of
\myeqref{eq:approx_volume_penalty} and controlled by the compression parameter
$\kappa_v$:
\begin{align}
  \mathbf{f}_v(\mathbf{q}) = -\frac{(V - V_0)}{V_0\kappa_v}\frac{\partial V}{\partial \mathbf{q}}. \label{eq:approx_volume_force}
\end{align}
This can then be added directly to \myeqref{eq:f}. Incidentally, the Jacobian of \myeqref{eq:approx_volume_force}
is dense, however, it can be approximated by the sparse term involving $\partial^2 V/\partial \mathbf{q}^2$, which
expresses only local force changes.  In matrix-free solvers where only matrix-vector products are required,
the complete derivative can be computed without hindering performance since the
full dense Jacobian is never stored in memory.

\section{Numerical Methods}
\label{sec:numerical_methods}

In this section we outline and motivate methods for approximately solving
the non-linear force balance system \eqref{eq:eom}.

\subsection{Time Integration}
\label{sec:time_integration}

The equations of motion \eqref{eq:eom} can be discretized in time using a variety of implicit
time integration methods.
Purely explicit integration schemes are not recommended since they prohibitively
restrict the admissible time step size in stiff problems --- contact and
friction can produce extremely large forces causing instability that is
intrinsic to the problem we are trying to solve.

Using standard notation, we assume that at time $t$ we know $\q = \q^t$ and $\vv=\vv^t$,
and employ a step size $h$ to proceed forward in time.
The integration methods we consider can be expressed by the momentum balance equation
\begin{align}
    \genv{0} = \rr(\vv^{\tt+h}; h, \f, \M), \label{eq:momentum_balance}
\end{align}
where we use superscripts to indicate time. Each integration scheme is characterized by
one or more residual functions $\rr$ used to determine the final velocity $\vv^{\tt + h}$.
Except for trapezoidal scheme, all integrators we consider are L-stable, indicating that
they dampen high frequency error components for stiff and highly oscillatory or unstable problems.
Elastodynamics with frictional contacts can exhibit instabilities
(see Section I.B. in the supplemental document), however, we know that
friction is naturally dissipative, and so we expect solutions to behave stably overall.
L-stability ensures that any additional stiffness present in
the system will not destabilize the numerical solution. For further
discussion on stability see \cite{ascher98}.

\subsubsection{Backward Euler}
The simplest implicit scheme is backward Euler (BE), defined by
\begin{align}
    \rr_{\BE}(\vv^{t+h}) & = \M(\vv^{\tt +h} - \vv^{\tt}) - h\f(\q^{\tt} + h\vv^{\tt+h},\vv^{\tt+h}) = \genv{0}. \label{eq:be_residual}
\end{align}

\subsubsection{Trapezoidal Rule}
A well-known method for mixing explicitly and implicitly determined forces on the system is
the trapezoidal rule (TR), defined by the momentum balance
\begin{align}
    \rr_{\TR}(\vv^{\tt+h})     & = \M(\vv^{\tt+h} - \vv^{\tt}) \notag \\
    &- \frac h2 \left(\f(\q^{\tt}, \vv^{\tt}) + \f(\q^{\tt+h}, \vv^{\tt+h})\right) = \genv{0}, \notag \\
    \text{where } \q^{\tt+h} & = \q^{\tt} + \frac h2 \left(\vv^{\tt} + \vv^{\tt+h}\right). \label{eq:tr_residual}
\end{align}
Notably, this method is equivalent to the popular implicit
Newmark-$\beta$ \cite{newmark59, hughes00, geradin93} method with $\beta = 1/4$ and $\gamma = 1/2$. The frictional
contact problem we address is not stable, and can generate large
stiffnesses for high elastic moduli, large deformations or due to contact and
friction.  The smoothed frictional contact problem
produces high frequency oscillations, which are exacerbated by TR, whereas
ideally we want these to be damped away.  See
Section I.B. in the supplemental document for a concrete example.
In spite of these flaws, TR is still used in practice, often decoupled from the
frictional contact problem \cite{li20}.
In \secref{sec:ball_in_box} we demonstrate
how properly coupling TR as defined in \myeqref{eq:tr_residual} can resolve some
instabilities in practice.

It is also straightforward to use higher-order L-stable methods within our formulation.
BDF2, TR-BDF2 and SDIRK2 are defined in the supplemental document; for further discussion
of these methods see \cite{loschner20, ascher21}.

\subsection{Damped Newton}\label{sec:damped_newton}

The momentum balance \myeqref{eq:momentum_balance} can be solved efficiently by
second-order root-finding methods like Newton's. In the absence of constraints, this can be
seen as a generalization of incremental potential optimization \cite{kane00}, where
the merit function is set to be an energy potential\footnote{
    While the original incremental potential is intended to be optimized over
    positions, the velocity derivatives of all
    integrators we consider are a constant
    multiple of positional derivatives.
    Thus optimizing over velocity here is equivalent to optimizing over positions.
} $W(\mathbf{v})$ such that
$\partial W(\mathbf{v}) / \partial \mathbf{v} = \mathbf{r}(\mathbf{v})$.
However, in that case to maintain a descent direction, $\partial \rr/ \partial \vv$ must be
appropriately modified to remain positive definite.

Since the presence of friction forces precludes a single potential $W$ for
minimization \cite{desaxce98}, many methods relying on incremental potentials
build special workarounds to solve for exact Coulomb-based friction, including staggered projections
\cite{kaufman08}, fixed-point methods \cite{erleben17} and lagged friction \cite{li20}.
Others employ non-smooth Newton to find roots of a proxy function
\cite{bertails11,daviet11,kaufman14}.
\new{We employ the penalty-based frictional contact approach promoted by
Geilinger et al. \shortcite{geilinger20}.
We extend their approach with additional terms of the Jacobian relating to
changes in the sliding basis to ensure a more accurate descent direction and
parity with the method in \secref{sec:inexact_newton}.  In addition to
contacts against static boundaries, we also demonstrate how this method
performs in full bidirectional contact between elastic solids.
Empirically we found that in many common cases
omitting sliding basis derivatives can speed up simulations, especially when
contact Jacobians are complex, however we expect convergence can suffer in cases
with complex deforming contact surfaces and large time steps.  We leave a
comprehensive convergence study to future work since it would require more
careful contact handling.}
For lower resolution examples we use the damped Newton algorithm as defined in \algref{alg:damped_newton},
where the problem Jacobian defined by $\mathbf{J} = \partial \mathbf{r}/\partial \mathbf{v}$ is
a square and potentially non-symmetric matrix (see Section I.A. in the supplemental document).
Assuming that $\mathbf{J}$ is boundedly invertible in the neighborhood of the root,
for a sufficiently good initial estimate,
damped Newton is guaranteed to converge\footnote{If $\mathbf{J}$ is also sufficiently regular then (undamped) Newton convergence is Q-quadratic.} \cite{nocedal06}.
While singular Jacobians can cause problems, in our experiments they are rare, and often can be eliminated by decreasing
the time step in dynamic simulations. Furthermore, in
Section I.A. of the supplemental document we show that our method does not introduce singularities
through coupling between elasticity, contact and friction on a single node.

\begin{algorithm}[ht]
    \caption{\sc DampedNewton}
    \label{alg:damped_newton}
    \DontPrintSemicolon
    % Input
    \KwIn{\begin{align*}
        k_{\text{max}} &\gets \text{maximum number of Newton iterations} \\
        \mathbf{v} &\gets \text{previous velocities}
    \end{align*}}
    % Output
    \KwOut{$\mathbf{v}_k \gets$ velocity for the next time step}

    $\mathbf{v}_0 \gets \mathbf{v}$ \tcc*[r]{Initialize velocity}

    \For{$k \gets 0$ \KwTo $k_{\text{max}}$}{
        \If{\sc ShouldStop$(\mathbf{r}(\mathbf{v}_{k}), \mathbf{v}_{k})$}{
            \KwSty{break}\;
        }
        $\mathbf{p}_k \gets -\mathbf{J}(\mathbf{v}_k)^{-1}\mathbf{r}(\mathbf{v}_k)$ \tcc*[r]{Set search direction} \label{algline:linear_solve}

        $\alpha \gets \text{\sc LineSearch}(\mathbf{v}_k, \mathbf{p}_k)$\;
        $\mathbf{v}_{k+1} \gets \mathbf{v}_k + \alpha\mathbf{p}_k$\;
    }
\end{algorithm}

\subsection{Inexact Damped Newton}\label{sec:inexact_newton}
For large scale problems, it is often preferable to use an iterative linear solver, which
can outperform a direct solver when degrees of freedom are sufficiently abundant.
We use \emph{inexact Newton} to closely couple the iterative solver with our
damped Newton's method.

Since friction forces produce a non-symmetric Jacobian,
we chose the biconjugate gradient stabilized (BiCGSTAB) algorithm \cite{vandervorst92}
to find Newton search directions $\p_k$. For Jacobian $\J_k = \partial \rr_k /
    \partial \vv$ with $\rr_k = \rr(\vv_k)$, the search direction is determined by
\begin{align*}
    \left\|\rr_k + \J_k \p_k\right\| \leq \sigma_k\|\rr_k\|,
\end{align*}
where $\sigma_k = \min(\|\rr_k\|^\varphi/\|\rr_{k-1}\|^\varphi, \sigma)$ and $\varphi = (1 + \sqrt{5})/2$ to maintain Q-quadratic convergence
\cite{eisenstat96}.

Using BiCGSTAB as the iterative solver additionally allows one to
use forward automatic differentiation to efficiently compute products $\mathbf{J}\mathbf{p}$ \newTvcg{\cite{griewank08}}.

The final inexact Newton algorithm is presented in \algref{alg:inexact_damped_newton}.% is augmented with the assisted line search.

\begin{algorithm}[h]
    \caption{\sc InexactDampedNewton}
    \label{alg:inexact_damped_newton}
    \DontPrintSemicolon
    % Input
    \KwIn{\begin{align*}
        k_{\text{max}} &\gets \text{maximum number of Newton iterations} \\
        \mathbf{v} &\gets \text{previous velocities} \\
        c_1 &\gets 10^{-4} \\
        \newTvcg{\sigma} &\newTvcg{\gets 0.01}\\
        \newTvcg{\rho} &\newTvcg{\gets 0.5}
    \end{align*}}
    % Output
    \KwOut{$\mathbf{v}_k \gets$ velocity for the next time step}

    $\mathbf{v}_0 \gets \mathbf{v}$ \tcc*[r]{Initialize velocity}

    \For{$k \gets 0$ \KwTo $k_{\text{max}}$}{
        \If{\sc ShouldStop$(\mathbf{r}(\mathbf{v}_{k}), \mathbf{v}_{k})$}{
            \KwSty{break}\;
        }
        $\sigma_k \gets \min(\|\rr_k\|^\varphi/\|\rr_{k-1}\|^\varphi, \sigma)$\;
        Find $\mathbf{p}_k$ such that $\left\|\rr_k + \J_k \p_k\right\| \leq \sigma_k\|\rr_k\|$\;

        $\alpha \gets 1$\;
        \tcc*[l]{Backtracking}
        \While{$\|\mathbf{r}(\mathbf{v} + \alpha \mathbf{p})\| > (1 - c_1\alpha(1 - \sigma_k))\|\mathbf{r}(\mathbf{v}_k)\|$}{
            $\alpha \gets \rho\alpha$\;
        }
        $\sigma_k \gets 1 - \alpha(1 - \sigma_k)$\;
        $\mathbf{v}_{k+1} \gets \mathbf{v}_k + \alpha\mathbf{p}_k$\;
    }
\end{algorithm}

\subsection{Contact}
To ensure that no penetrations remain (i.e. $d_i > 0$ for each contact $i$) at the end of the time step we measure the deepest penetration
depth $\dd_{\text{deepest}} = \min_i(\dd_i(\q^{\tt+h}))$, and bump the contact stiffness parameter $\kappa$ by the factor 
$\frac{b'(d_{\text{deepest}})}{b'(0.5\delta)}$ whenever $d_{\text{deepest}} < 0$, where $b'$ is the scalar derivative of $b$. The
same step is then repeated with the new $\kappa$. This scheme sets the optimal
contact penalty value found by the Newton scheme to appear for contacts $0.5\delta$ outside
the contact surface. As such, in most cases one time step with an additional contact iteration
is sufficient before all contacts are resolved. Furthermore, $\kappa$ is never
decreased so long as there are active contacts to avoid oscillations at the
contact surface.

The downside of this technique is that it compromises the smoothness of the
simulation.  We postulate that in practice, this may not be problematic in
a differentiable pipeline since $\kappa$ is not changed frequently and
subsequent differentiable iterations can carry forward the maximal $\kappa$ to
maintain smoothness.

\subsection{\newTvcg{Correct TR integration in IPC}} \label{sec:ipccompat}
Li et al. \shortcite{li20} introduced incremental potential contact (IPC),
a robust method to resolve contacts by minimizing incremental potentials with friction being
evaluated using lagged positional estimates from the previous time step.
\new{Here we propose a simple fix for handling higher-order time integrators
in the IPC framework, and show how our formulation relates to this method.}
With the lagged friction approach, the BE and TR discretizations are given by 
\begin{align}
    \mathbf{r}_{\text{BE,IPC}}(\mathbf{v}^{t+h}) &= \mathbf{v}^{t+h} - \mathbf{v}^t  
     - h \mathbf{M}^{-1}\mathbf{f}_{\mathit{lag}}(\mathbf{q}^t, \mathbf{q}^{t+h}, \mathbf{v}^{t+h}), \label{eq:be_ipc} \\
    \mathbf{r}_{\text{TR,IPC}}(\mathbf{v}^{t+h}) &= \mathbf{v}^{t+h} - \mathbf{v}^t \notag \\
     - \frac{h}{2} \mathbf{M}^{-1}&\left(
         \mathbf{f}_{\mathit{lag}}(\mathbf{q}^{t}, \mathbf{q}^{t+h}, \mathbf{v}^{t+h})
         + \mathbf{f}_{\mathit{lag}}(\mathbf{q}^{t}, \mathbf{q}^{t}, \mathbf{v}^{t})
         \right), \label{eq:tr_ipc}
\end{align}
respectively, where we write
\begin{align}
     \mathbf{f}_{\mathit{lag}}(\mathbf{q}^t, \mathbf{q}^{t+h}, \mathbf{v}^{t+h})
     = \mathbf{f}_{\mathit{edcg}}(\mathbf{q}^{t+h}, \mathbf{v}^{t+h}) + \mathbf{f}_f(\mathbf{q}^t, \mathbf{v}^{t+h}).
\end{align}
Here $\mathbf{f}_{\mathit{edcg}}$ is the sum of elastic, damping,
contact and external forces and $\mathbf{f}_f$ is the friction
force as before.
\new{Notice that in \eqref{eq:tr_ipc} the entire net force $\mathbf{f}_{\mathit{lag}}$
is split into implicit and explicit parts, whereas the original proposal
\cite{li20} for TR in IPC is to apply this splitting to non-contact forces only.
In \secref{sec:ball_in_box} we show how solving \eqref{eq:tr_ipc} will generate more
stable results when compared to the original IPC implementation.}

\new{Note that $\mathbf{f}_{\mathit{lag}}$ has a well-defined antiderivative with respect to
$\mathbf{v}^{t+h}$, which can be minimized using common optimization tools.}
In this view, IPC effectively solves Eqs. \eqref{eq:be_ipc} or \eqref{eq:tr_ipc}
using the proposed log-barrier potential as a merit function, CCD
aided line search and Hessian projection.
Although this can be done iteratively with better estimates for the lagged friction force,
this approach has limitations as demonstrated in \secref{sec:blockslide}.

\section{Results} \label{sec:results}
All experiments were run on the AMD Ryzen Threadripper 1920X CPU with 12 cores,
24 threads at 3.7 GHz boost clock and 32 GB RAM. We used Blender 3.1
\shortcite{blender21} for all 3D renderings.
For \algref{alg:damped_newton} we used the Intel MKL sparse LU solver to solve
the square non-symmetric linear system on line~\ref{algline:linear_solve}.
We used dual numbers for forward automatic differentiation 
\cite{larionov22autodiff} and a custom BiCGSTAB implementation for the inexact Newton Algorithm~\ref{alg:inexact_damped_newton}.
In the following results Algorithms~\ref{alg:damped_newton} and \ref{alg:inexact_damped_newton}
are dubbed ``Direct'' and ``Iterative'', respectively, since the former uses a direct
linear solver and the latter uses an iterative linear solver.
\new{We did not evaluate viscous friction effects here, so $\mu_v = 0$ in all examples.
For examples using the same dynamic and static friction coefficients $\mu_d$ and $\mu_s$, we omit the subscript.}

\subsection{Friction accuracy} \label{sec:friction_accuracy}
With the following examples we demonstrate two scenarios where lagged friction causes large deviations from an
expected accurate and stable friction response.

\subsubsection{Block slide} \label{sec:blockslide}

In this example we let a stiff elastic block slide down a 10 degree slope
expecting it to stop for $\mu = 0.177 > \tan(10^\circ)$ after sliding for
a total of $x_T = 0.769$ m for $T = 15.38$ seconds (see supplemental document for details on the experiment).

\figref{fig:blockslide} demonstrates that our method produces consistent
stopping across a variety of time step sizes using BE and TR time integration.
We compare against IPC \cite{li20}, a state-of-the-art smoothed friction method
using a lagged friction approach to show that it fails to establish consistent
stopping with BE, and fails to stop with TR altogether after 50 seconds. We
reproduce the lagged friction method in our simulator
to demonstrate that TR can be used to generate reliable stopping if the equations of motion are
correctly integrated as in \myeqref{eq:tr_ipc}.
Our method produces a more accurate stopping distance using TR than IPC does using
BE even after using multiple fixed point iterations.

\begin{figure*}
    \centering
    \begin{subfigure}{\linewidth}
        \includegraphics[width=\linewidth]{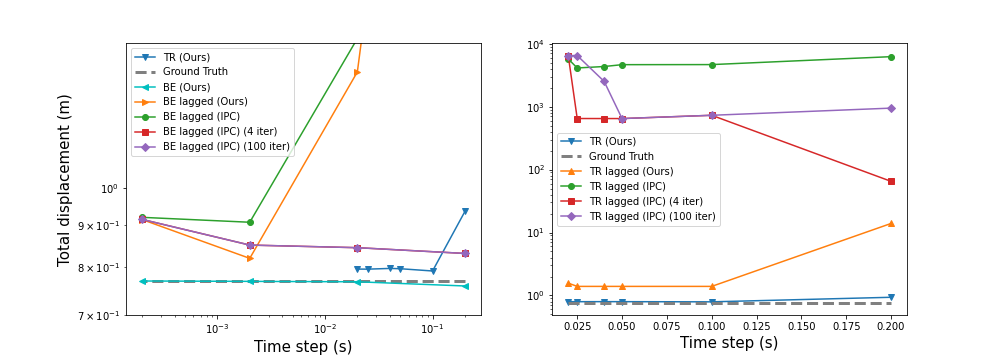}%
        \caption{ Total displacement travelled by the block before coming to a
        stop. With BE and multiple fixed point iterations the lagged friction
        approach can produce a reasonable approximation of the stopping
        behavior, however, the approximation does not converge to the true
        solution. Our method produces an accurate estimate of the true behavior
        with BE, while TR produces a better estimate than lagged friction at $h
        = 0.1$ s and lower.  With lagged friction, the TR method presented in
        \myeqref{eq:tr_ipc} produces a more accurate result than IPC since contact
        is handled together with other implicit forces.
        \label{fig:blockslide_dist}}
    \end{subfigure}
    \begin{subfigure}{\linewidth}
        \includegraphics[width=\linewidth]{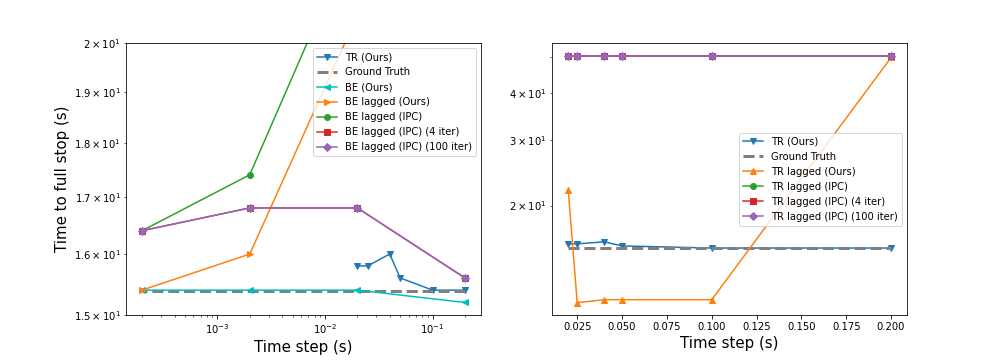}%
        \caption{Time taken by each block to come to a stop.
        With BE, our method produces an accurate approximation to the true stopping time for all but
        the largest time step. In contrast, lagged friction fails to converge to the true stopping time
        under multiple fixed point iterations. With TR, the approximation is not as accurate as with BE,
        although still closer to the true value than lagged friction using BE. With TR, lagged friction does not
        produce a reliable stopping time, even under refinement. Using IPC's TR
        implementation, the box does not stop after 50 s for any time step.
        \label{fig:blockslide_time}}
    \end{subfigure}
    \caption{\emph{Block slide.} Comparisons of analytic stopping conditions of a sliding block to numerical results.\label{fig:blockslide}}
\end{figure*}

\subsubsection{Bowl grasp}
\label{sec:bowl_grasp}

Control over the friction coefficient is particularly important in grasping scenarios since grasped objects are often
delicate. This means that friction forces involved in lifting are often close to the sliding threshold.

\begin{figure}
        \centering
        \includegraphics[width=0.8\linewidth]{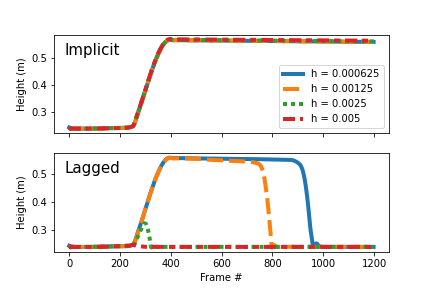}%
        \caption{The height of the bowl in~\figref{fig:bowl_grasp} is plotted against
        frame number as time step $h$ is varied.
        Here the bowl slips for the lagged friction method,
        whereas the fully implicit method maintains stable sticking for every
        time step, hence all plotted lines overlap.\label{fig:bowl_grasp_height}}
\end{figure}

As shown in~\figref{fig:bowl_grasp}, an upside down bowl is lifted using 3 soft pads to compare sticking stability of
\emph{lagged} friction given in \myeqref{eq:be_ipc} against a fully implicit method from \myeqref{eq:be_residual}.
The bowl is successfully picked up and stuck to the pads for a range of time steps when using the \emph{implicit} method, however
it slips for different time step values with \emph{lagged} friction. The height of the bowl is plotted in \figref{fig:bowl_grasp_height} for
each method and time step combination.

\subsubsection{Ball in a box}
\label{sec:ball_in_box}

\begin{figure}[h]
    \begin{subfigure}{\linewidth}
    \centering
    \includegraphics[width=0.66\linewidth]{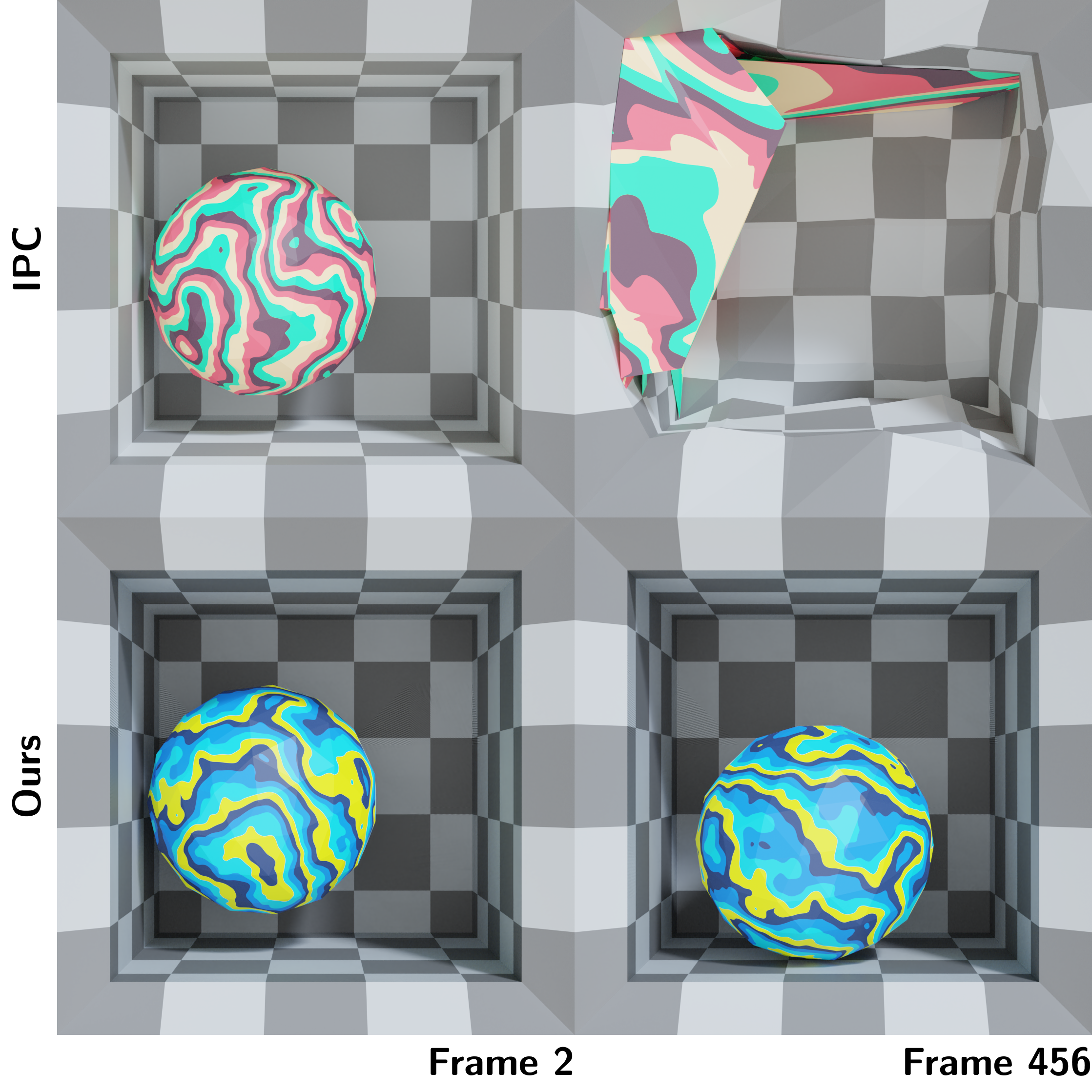}%
    \caption{The TR formulation proposed in IPC is unstable (top row) whereas our
    method produces a dissipative scenario (bottom row) even with a slightly weaker
    damping.\label{fig:ball_in_box_ipc_tr_compare_frames}}
    \end{subfigure}
    \begin{subfigure}{\linewidth}
    \centering
    \includegraphics[width=\linewidth]{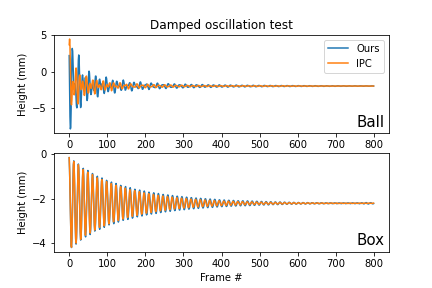}
    \caption{The ball (top) and the box (bottom) are simulated individually with both methods and centroid heights are plotted
    to ensure that the oscillation amplitudes with TR as implemented by IPC do not exceed those generated by our implementation.  \newTvcg{This test
    ensures that the reason for blowup in IPC is not due to integration of elasticity equations alone, but indeed due to loose coupling between
    elastic and contact terms.} \label{fig:ball_in_box_damping_test}}
    \end{subfigure}
    \caption{\emph{Ball in a box.} A spinning ball bounces inside an elastic box. Here $\rho = 10000$ kg/m${}^3$,
    $E = 500$ KPa, and $\nu = 0.1$ for the box and $\rho = 500$ kg/m${}^3$, $E = 50$ KPa and $\nu = 0.45$ for the ball,
    with friction coefficient set to $\mu = 0.1$ between them.
    For IPC, damping ratio is set to 0.02, and for our model, the damping parameter is set to 0.1 Hz. In both cases
    Rayleigh damping is used. 
    \label{fig:ball_in_box_ipc_tr_compare}}
    \Description{}
\end{figure}

A rubber ball placed inside an elastic box has an initial spin of 4800 rotations
per minute and an initial velocity set to $\vec{v}_0 = (-0.923, -0.385, 0)$.
In \figref{fig:ball_in_box_ipc_tr_compare},
we demonstrate how our fully coupled TR integrator produces more stable dynamic simulations
compared to the decoupled TR as proposed by Li et al. \shortcite{li20}.
This scenario is simulated with both methods for 800 frames at $h = 0.01$ s with
comparable damping parameters as shown in \figref{fig:ball_in_box_damping_test}.
The TR implementation used by IPC blows up, whereas in our formulation the energy is
eventually dissipated as expected.

\subsection{Performance}
In this section, we show how various combinations of volume preservation and
frictional contact constraints can affect the performance of the simulation. In
addition, we compare Algorithms~\ref{alg:damped_newton} and \ref{alg:inexact_damped_newton}
in performance and memory usage.

\subsubsection{Tube cloth bend}\label{sec:tube_cloth_bend}
\begin{figure}[h]
    \centering
    \begin{overpic}[width=0.33\linewidth,trim=280 0 400 0,clip]{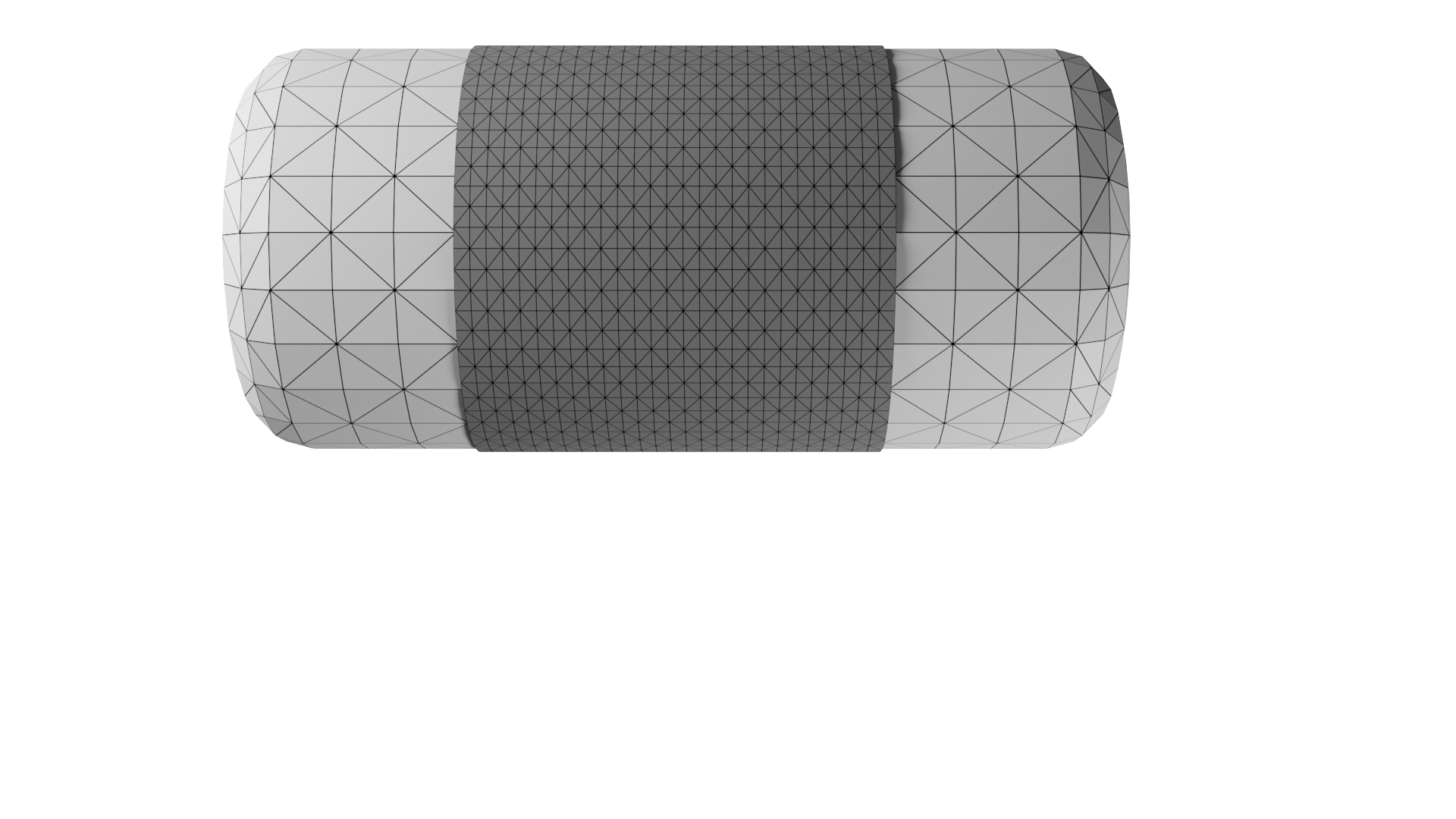}
        \put (5,17) {3K Triangles}
        \put (5,5)  {5K Tetrahedra}
    \end{overpic}%
    \includegraphics[width=0.33\linewidth,trim=280 0 400 0,clip]{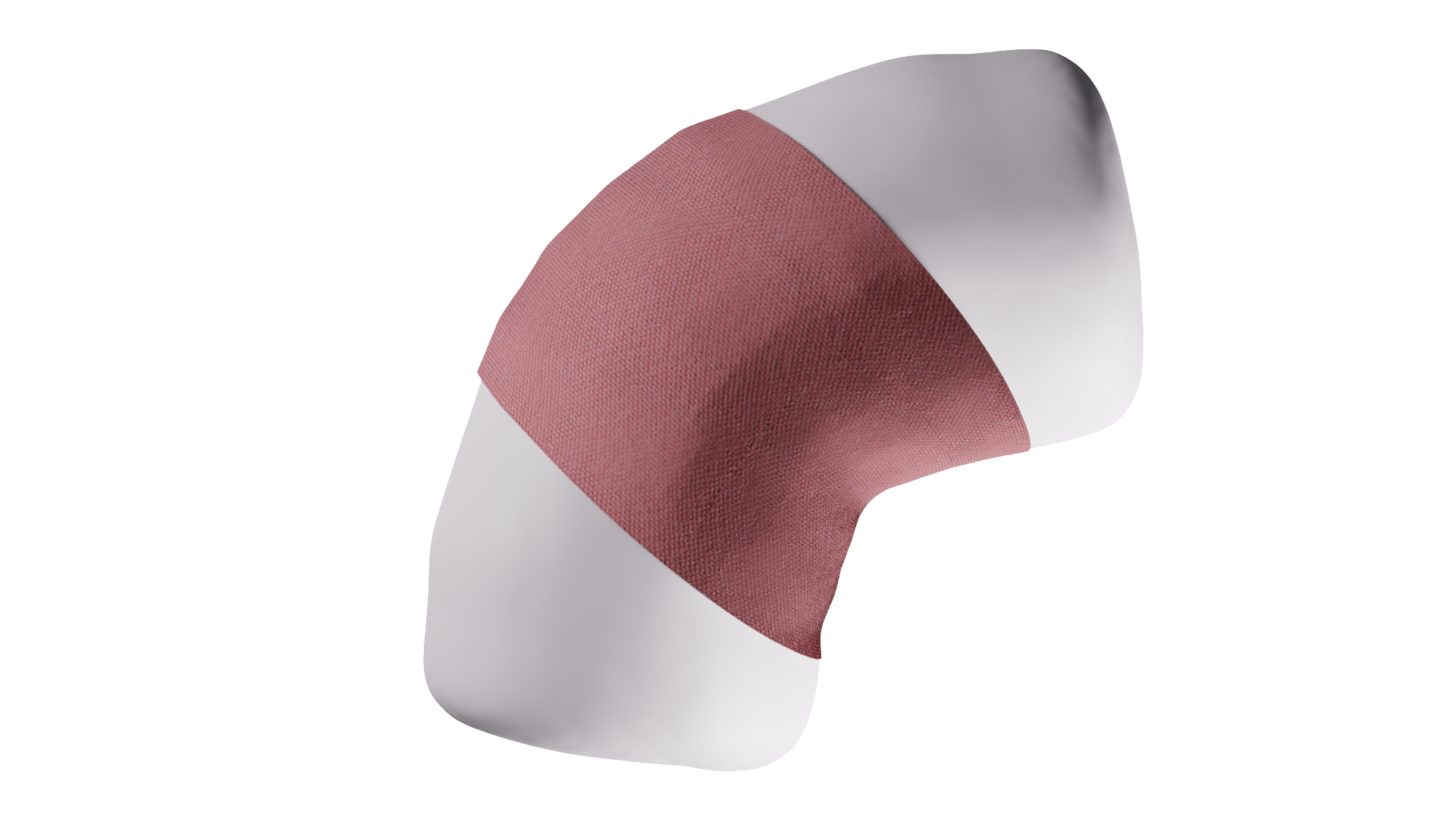}%
    \includegraphics[width=0.33\linewidth,trim=280 0 400 0,clip]{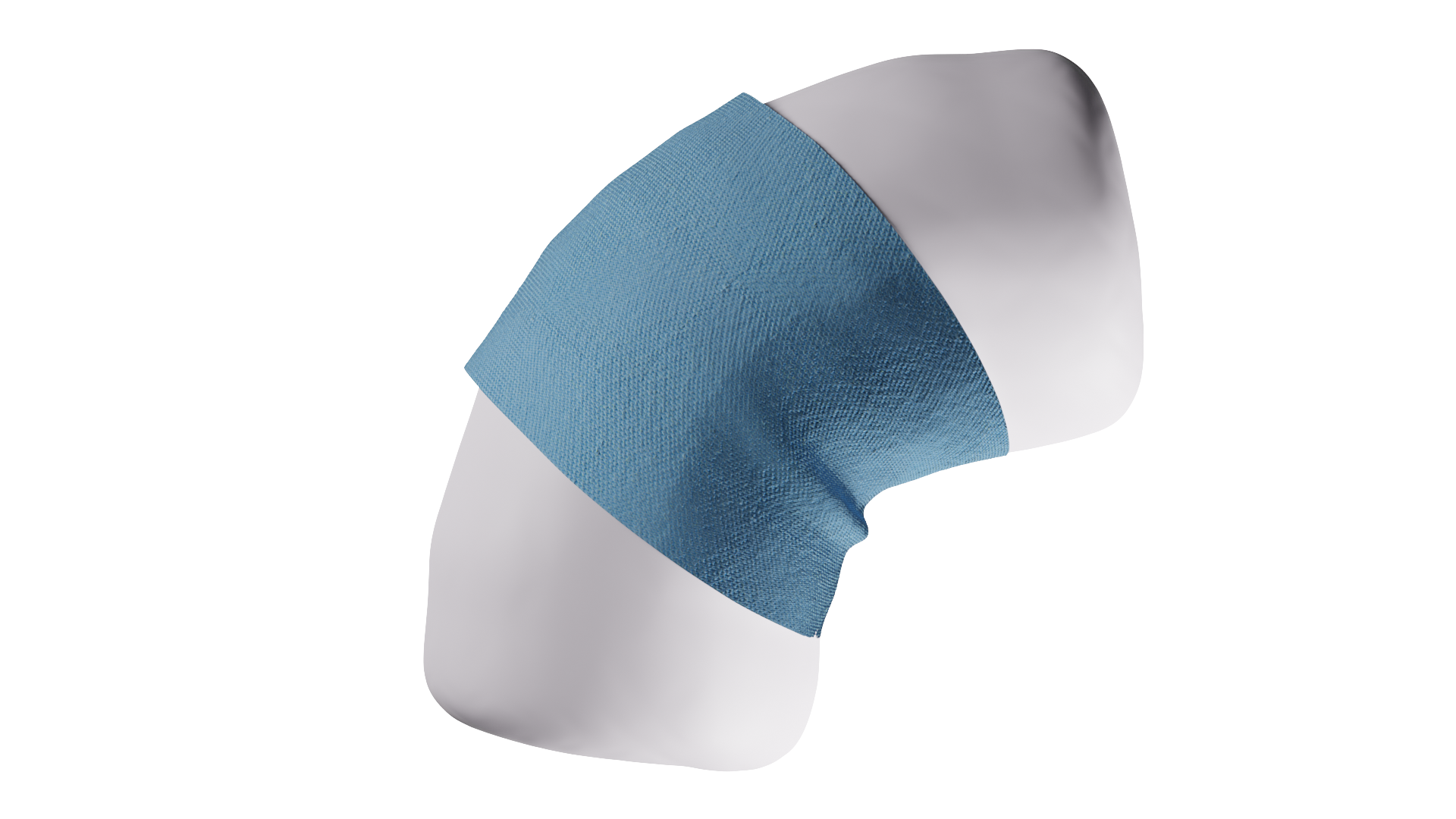}
    \begin{overpic}[width=0.33\linewidth,trim=280 0 400 0,clip]{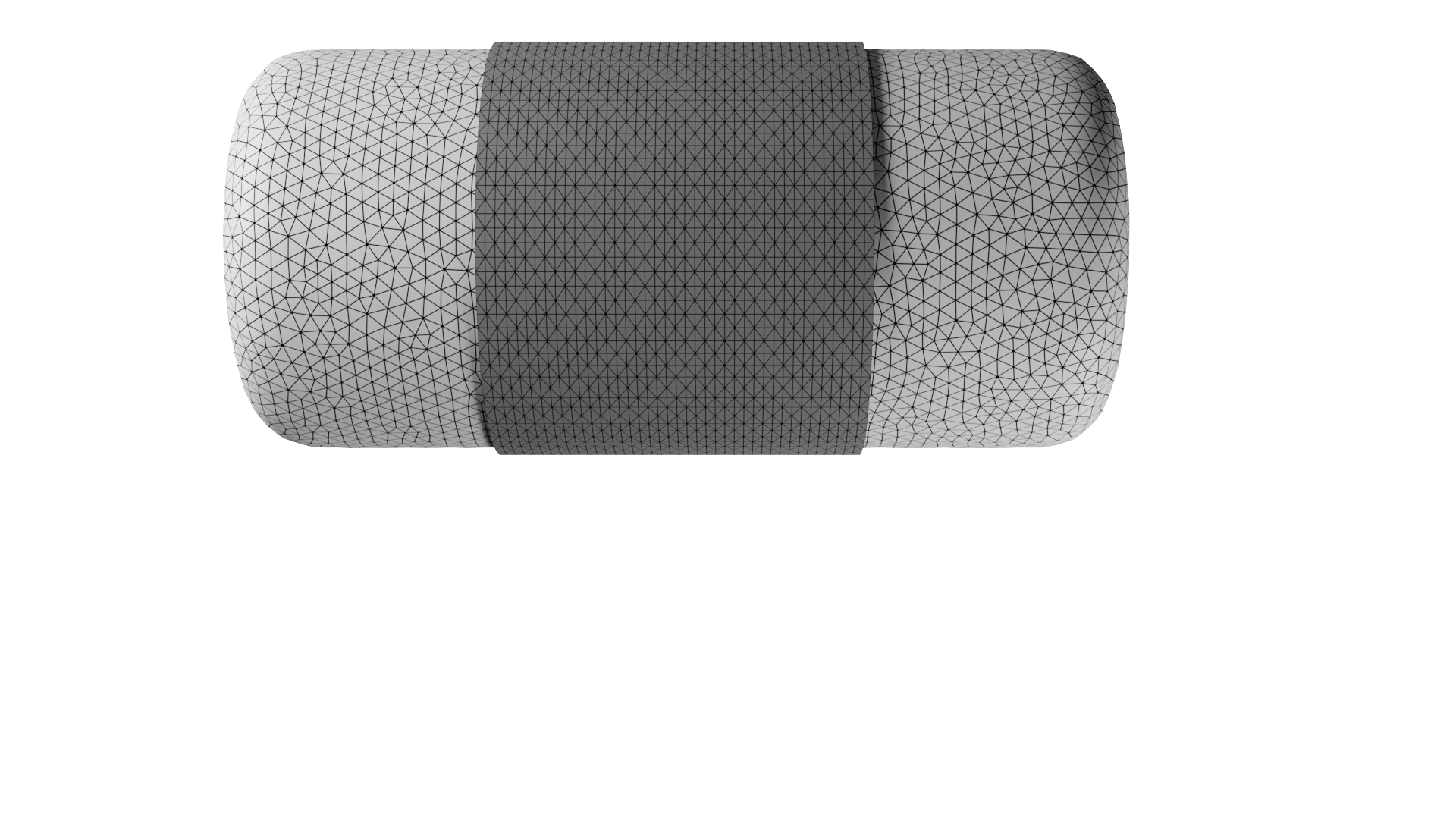}
        \put (5,17)  {8K Triangles}
        \put (5,5)  {30K Tetrahedra}
    \end{overpic}%
    \includegraphics[width=0.33\linewidth,trim=280 0 400 0,clip]{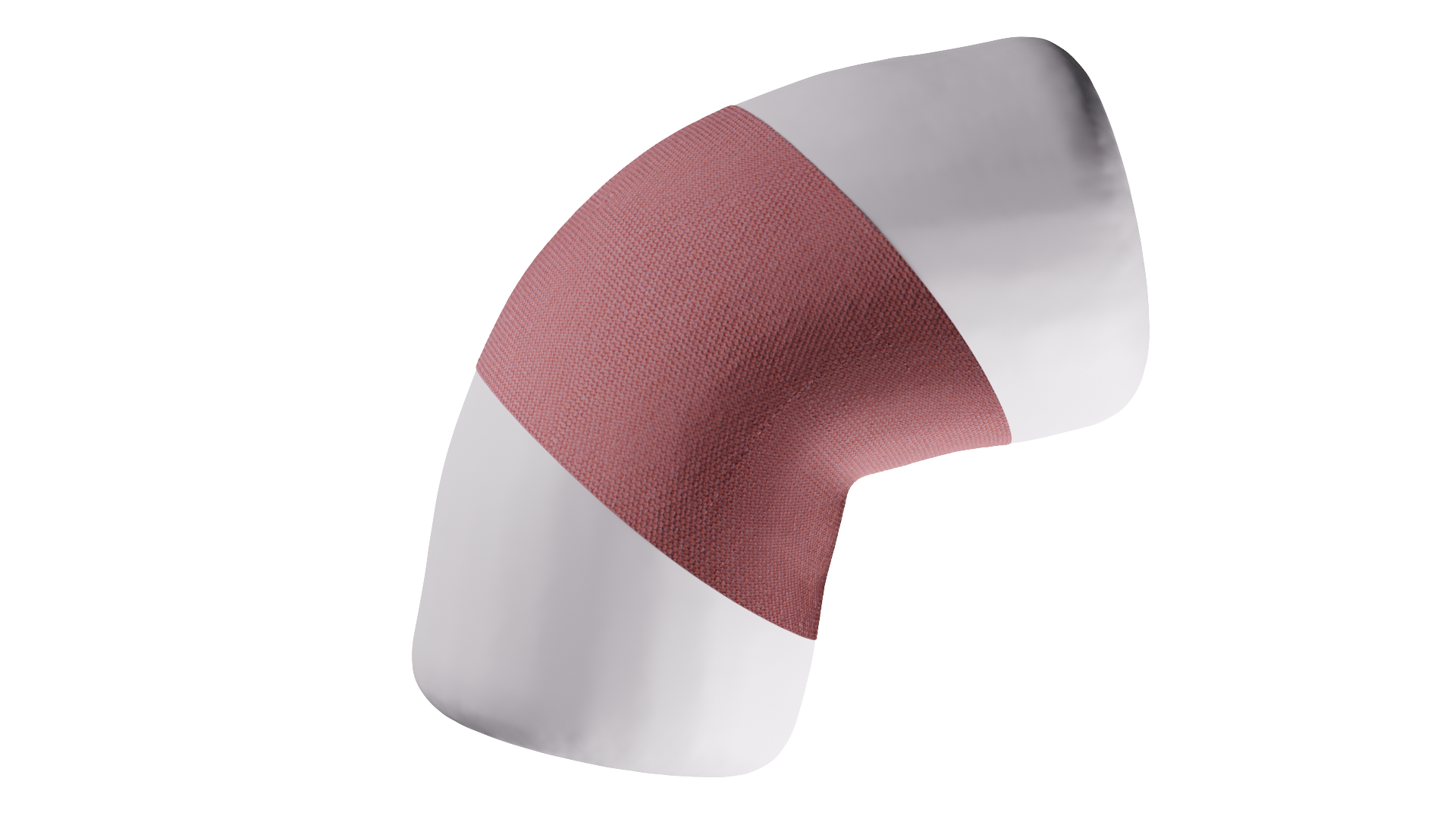}%
    \includegraphics[width=0.33\linewidth,trim=280 0 400 0,clip]{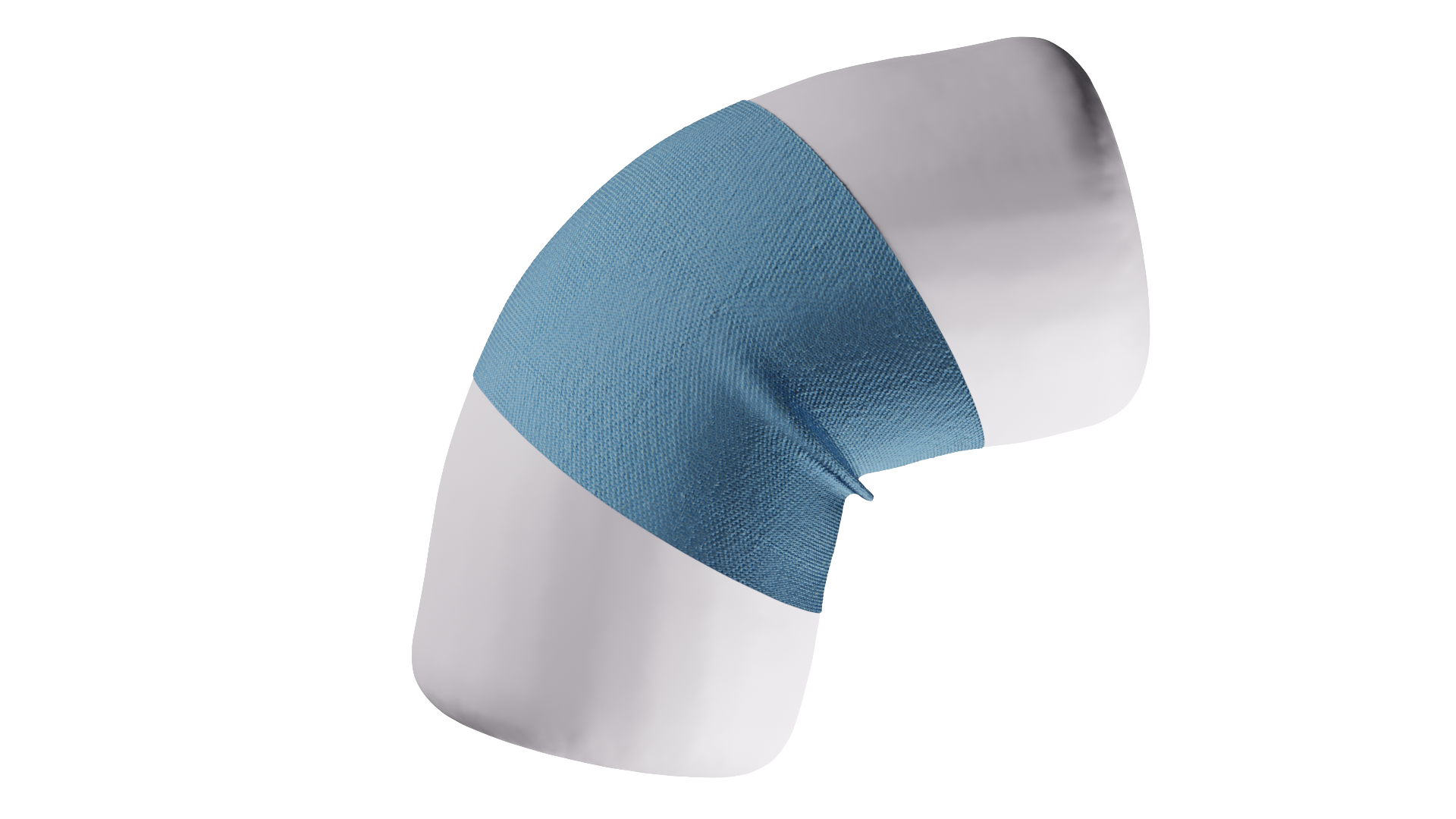}%
    \caption{\emph{Tube cloth bend.} A cylindrical garment is wrapped around a bending capsule.
    Lower resolution example is shown at the top row and higher resolution at the bottom row.
    The first column shows the initial configuration, second column shows the end result with low friction ($\mu = 0.2$),
    and the last column shows the end result with high friction ($\mu = 0.8$).
    \label{fig:bend}} 
\end{figure}

The inexact Newton Algorithm~\ref{alg:inexact_damped_newton} shines particularly
in scenarios with numerous contacts, such as with tight-fitting
garments, where the entire garment is in contact with a body.  Here we simulate
a simplified scenario of a tube cloth wrapped around a bending soft object
resembling an elbow or knee as depicted in~\figref{fig:bend}.  \tabref{tab:bend}
shows the corresponding timing results, which indicate that inexact Newton
performs much better than the damped Newton algorithm employing a direct solver.
Furthermore, the performance gap becomes large when the number of elements is increased.
\newTvcg{The higher resolution example in \tabref{tab:bend} does not include the ``Direct'' solver
because it is intractable at that resolution due to the scalability limitations of the direct solve.
In this case exact Newton can use an iterative solver, however a full comparison of the trade-offs
between different solvers is outside the scope of the paper since this type of solver would
require an additional fixed tolerance, which would affect performance and convergence characteristics.}
Interestingly this data also indicates that larger friction coefficients cause a
bigger bottleneck for the solve compared even to stiff volume change penalties
(indicated by small $\kappa_v$).

\begin{table}[h]
    \footnotesize
    \begin{center}
    \setlength{\tabcolsep}{4pt}
    \begin{tabular}{cccccccc}
        \toprule
        \# Elements &
        $\mu$ &
        Solver Type &
        $\kappa_v$ &
        Time &
        Memory &
        Volume Loss \\
        \midrule
        \multirow{4}*{
            \begin{tabular}{c}
            3K Tris \\
            5K Tets
            \end{tabular}
        } &
        \multirow{2}*{0.2} &
        % 1h20m08s
        Iterative &  N/A  & 4.81 & 1.84 GB & 0.962\% \\
        % 6h08m15s
        && Direct &  N/A  & 22.1 & 6.51 GB & 0.962\%\\ \cmidrule{2-7}
        &\multirow{2}*{0.8} &
        % 3h32m04s
        Iterative &  N/A  & 12.7 & 1.91 GB & 0.962\%\\
        % 7h06m35s
        && Direct &  N/A  & 25.6 & 7.09 GB  & 0.962\%\\
        \midrule
        \multirow{4}*{
            \begin{tabular}{c}
            8K Tris \\
            30K Tets
            \end{tabular}
        } &
        \multirow{2}*{0.2} &
        % 2h23m05s
        Iterative    &  N/A     & 8.59 & 830 MB & 2.55\% \\
        % 3h23m01s
        && Iterative & 4.6e-5 & 12.2 & 858 MB  & 2.44e-4\% \\ \cmidrule{2-7}
        &\multirow{2}*{0.8} &
        % 14h19m44s
        Iterative    &  N/A   & 51.6  & 625 MB & 2.57\% \\
        % 14h46n59s
        && Iterative & 4.6e-5 & 53.2  & 918 MB & 2.44e-4\%\\
        \bottomrule
    \end{tabular}
    \caption{\emph{Tube cloth bend performance data.} Time is measured in
    seconds per frame and memory refers to total memory growth during the
    simulation as reported by Houdini's performance monitor. Simulations without
    volume change penalty have no $\kappa_v$.}
    \label{tab:bend}
    \end{center}
\end{table}

\subsubsection{Ball Squish}
\begin{figure}
    \centering
    \includegraphics[width=0.9\linewidth,trim=0 90 0 0]{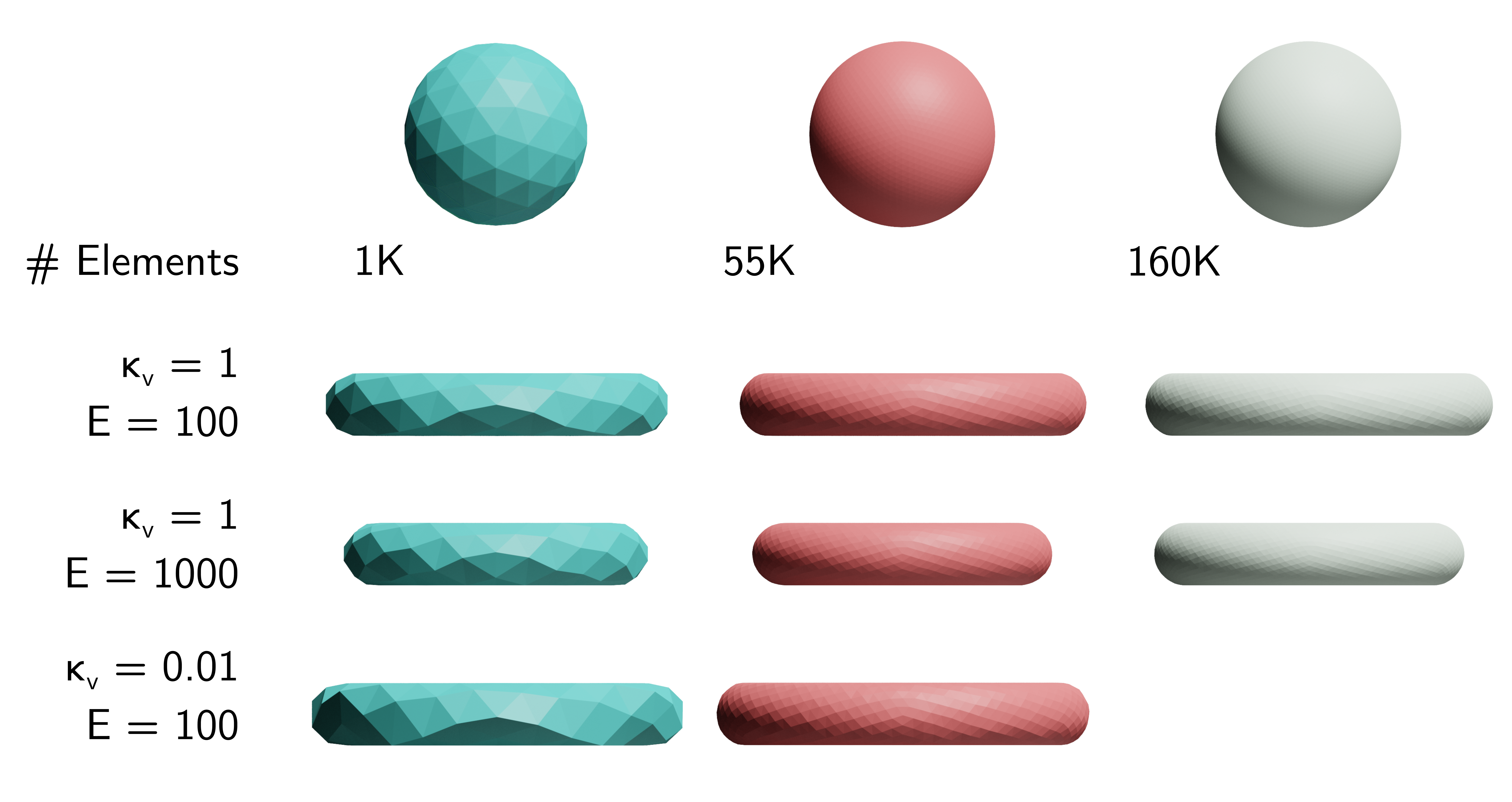}
    \caption{\emph{Ball squish.} A hollow ball of various resolutions is squished between two parallel rigid plates.
    The same simulations are performed for varying Young's moduli $E$ and compression coefficients $\kappa_v$.
    As expected, the squished ball occupies a larger volume for smaller $\kappa_v$ (last row), and, possibly due to locking artifacts,
    occupies a smaller volume for larger $E$.
    \label{fig:ball_squish}} 
\end{figure}

A hollow ball at various resolutions (1K, 55K and 160K elements) is pressed between two flat planes
as demonstrated in \figref{fig:ball_squish}.
As a result the ball experiences volume loss. To preserve some of the volume we
simulate the compression with compression coefficients $\kappa_v = 1$
(e.g. a ball filled with air) and $\kappa_v = 0.01$ (e.g. a ball filled with water).
In the latter case we expect significantly less volume loss, which is reflected in our experiments as shown
in \tabref{tab:ball_squish_timings}.  Furthermore, we note from
\tabref{tab:ball_squish_timings_incompressibility} that scenarios with small
$\kappa_v$ favor the ``Iterative'' method.  In \tabref{tab:ball_squish_timings_hessian_approx}
we see that this is true whether $\Jac$ is sparsely approximated (``Inexact'') or not (``Exact'').
In contrast, stiffer scenarios prefer the ``Direct'' method due to worse system conditioning.

% Modes: iterative, direct
% Volume coefficients: 1, 0.01, none
% Hessian approximation: true, false
% Resolution 1K, 55K, 160K
% Stiffness: 1000, 100

% Here we compare Direct vs Iterative methods in execution time and memory usage.
% We want to show 4 criteria in which iterative methods perform well:
% 1. Large resolution meshes
% 2. a. Small compressibility coefficients (Nearly incompressible volumes)
%    b. Exact volume constraint Hessians in direct methods.
% 3. Soft configurations

\begin{table}[ht]
\begin{subtable}[h]{\linewidth}
    \centering
    \begin{tabular}{l|ccc}
    \toprule
    \# Elements & 1K          & 55K        & 160K      \\
    \midrule
    Direct      & 0.354 (0.228) & 15.9 (1.69)  & 61.0 (4.26) \\
    Iterative   & 0.391 (0.081) & 31.9 (0.719) & 111 (1.92)  \\
    \midrule
    Volume loss & 33.6\%      & 27.8\%     & 27.7\% \\
    \bottomrule
    \end{tabular}
    \caption{$E = 100$ KPa, $\kappa_v = 1$. \label{tab:ball_squish_timings_resolution}}
\end{subtable}

\begin{subtable}[h]{\linewidth}
    \centering
    \begin{tabular}{l|cc}
    \toprule
    \# Elements & 1K         & 55K        \\
    \midrule
    Direct      & 1.08 (0.478)  & 89.1 (2.03)  \\
    Iterative   & 0.463 (0.214) & 42.6 (0.884) \\
    \midrule
    Volume loss & 0.49\%      & 0.36\% \\
    \bottomrule
    \end{tabular}
    \caption{$E = 100$ KPa, $\kappa_v = 0.01$. \label{tab:ball_squish_timings_incompressibility}}
\end{subtable}

\begin{subtable}[h]{\linewidth}
    \centering
    \begin{tabular}{l|ccc}
    \toprule
    $\kappa_v =$              & 1           & 0.01        \\
    \midrule
    Direct Exact $\Jac$       & 0.832 (0.393) & 0.871 (0.672) \\
    Direct Inexact $\Jac$     & 0.354 (0.228) & 1.08  (0.478) \\
    Iterative                 & 0.391 (0.081) & 0.463 (0.214) \\
    \bottomrule
    \end{tabular}
    \caption{$E = 100$ KPa, for 1K elements. \label{tab:ball_squish_timings_hessian_approx}}
\end{subtable}

\begin{subtable}[h]{\linewidth}
    \centering
    \begin{tabular}{l|ccc}
    \toprule
    \# Elements & 1K          & 55K        & 160K       \\
    \midrule
    Direct      & 0.312 (0.388) & 8.28 (1.52)  & 25.5 (4.14)  \\
    Iterative   & 0.593 (0.097) & 28.3 (0.732) & 56.2 (2.62) \\
    \midrule
    Volume loss & 74.3\%      & 74.6\%     & 75.0\% \\
    \bottomrule
    \end{tabular}
    \caption{$E = 1000$ KPa with no volume preservation constraint. \label{tab:ball_squish_timings_stiffness}}
\end{subtable}

\caption[Performance data for the ball squish example]{
\emph{Ball squish timings, memory usage (in parentheses) and volume loss.}
A hollow ball is squished between two flat rigid plates at time step $h=0.001$ s.
Timings are given in seconds per frame and averaged over 804
frames. The memory usage measured in GB over the entire simulation sequence is shown in parentheses.
Volume loss is computed for the change in volume between frames 1 and 804 as a percentage of initial
volume inside the ball. Each table specifies the number of elements, Young's modulus $E$ and compression coefficient
$\kappa_v$ where applicable. Here we compare how the performance characteristics of our
``Direct'' and ``Iterative'' methods change when problem stiffness and
$\kappa_v$ are varied for different mesh resolutions.
\label{tab:ball_squish_timings}}
\end{table}

\subsection{Real world phenomena}
In the following example we show how our simulator can reproduce deformations captured in the real world.

\subsubsection{Tennis ball}\label{sec:tennis_ball}
\begin{figure}
    \centering
    \includegraphics[width=0.24\linewidth,trim=400 200 400 0,clip]{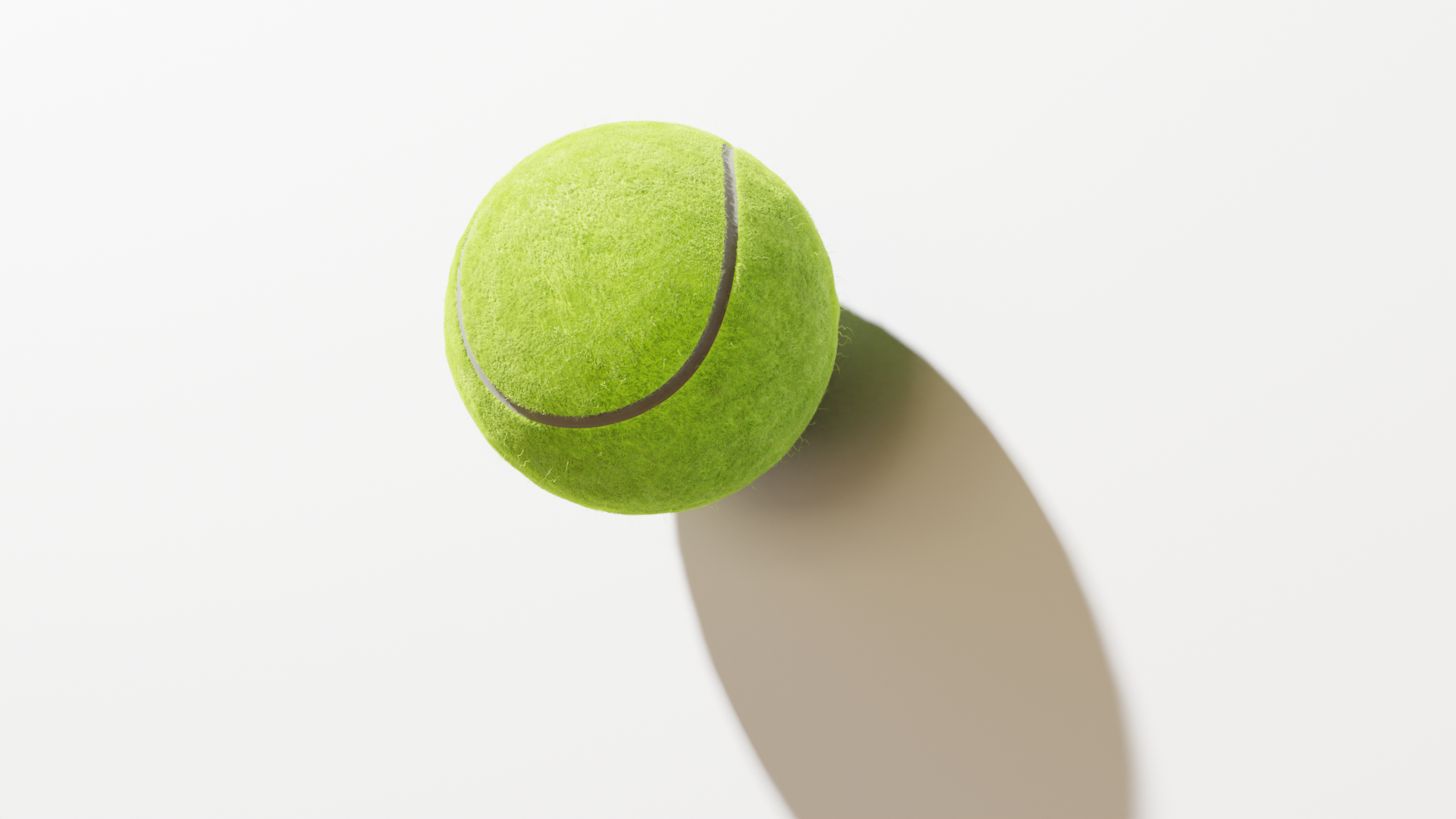}%
    \includegraphics[width=0.24\linewidth,trim=400 200 400 0,clip]{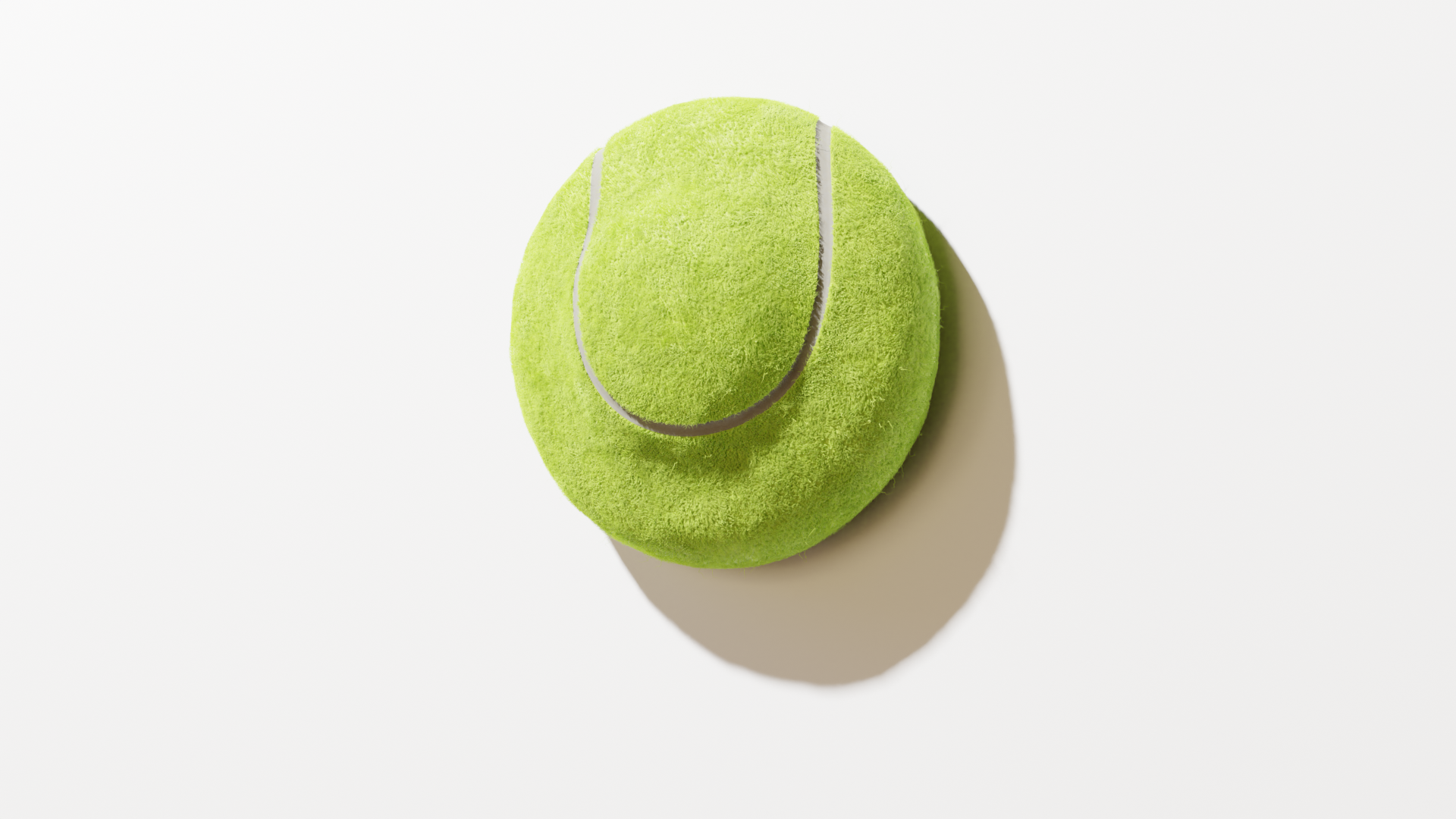}%
    \includegraphics[width=0.24\linewidth,trim=400 200 400 0,clip]{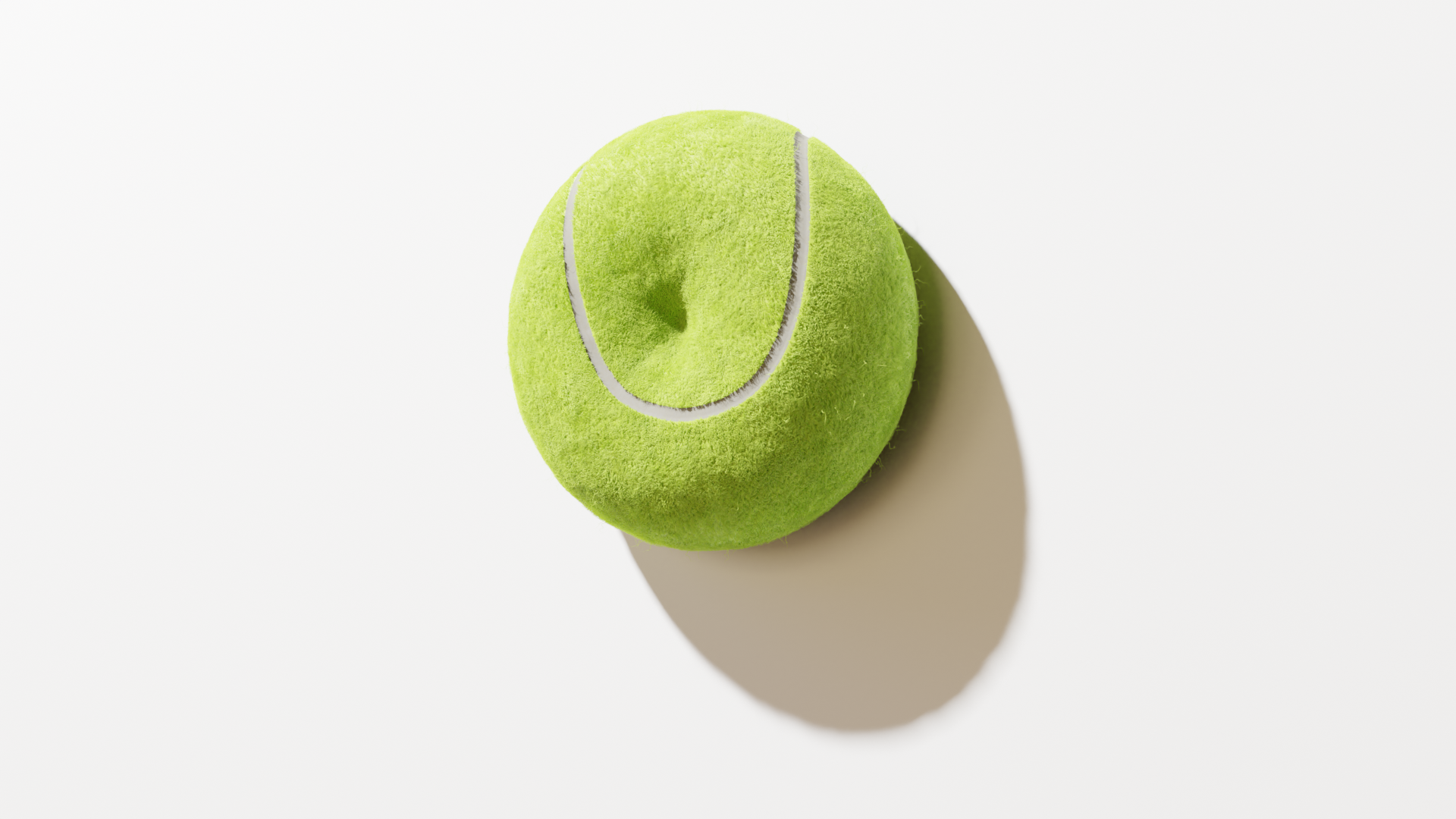}%
    \includegraphics[width=0.24\linewidth,trim=400 200 400 0,clip]{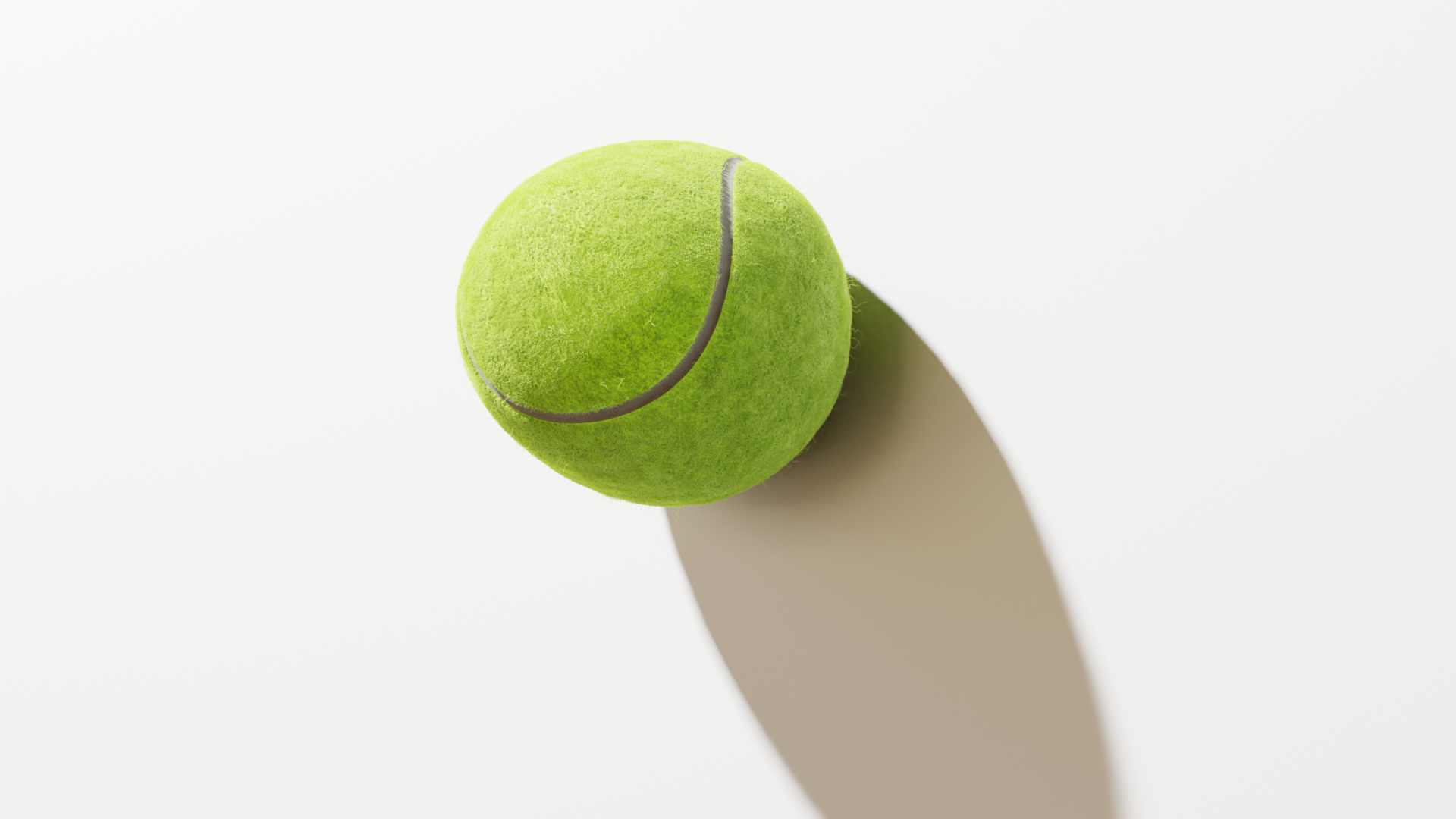}
    \caption{\emph{Highspeed tennis ball collision.} Simulated hollow tennis
    ball collides against a wall at 100mph.  The deformation here
    closely resembles that of a real tennis ball as captured by
    Anderson \shortcite{anderson18}\label{fig:ball_100mph}.}
\end{figure}
Tennis ball dynamics is a prime example for all methods proposed in this paper.
\newTvcg{Volume preservation and higher-order integration enables accurate bounce behavior, whereas
accurate friction is needed for predicting the bounce trajectory. Here we demonstrate the need
for the former.}

First, we launch a tennis ball at a wall at 100 mph (44.7 m/s) to reproduce accurate
slow motion deformation. Tennis balls are typically
pressurized to approximately 1 atm above atmospheric pressure to
produce a livelier bounce during play. In \figref{fig:ball_100mph}
we show how deformation changes when the ball is pressurized and compare the result
with live footage. The simulation contains 100K tetrahedra and 92K vertices. The ball
is hollow with a stiff inner \newTvcg{layer} ($E = 6$ MPa, $\nu = 0.4995$ and $\rho = 934$ kg/m${}^3$)
and a light outer felt material ($E = 5.4$ MPa, $\nu = 0.3$ and $\rho = 4.69$ kg/m${}^3$).
The volume change penalty is applied to the interior of the ball. This 1000 frame simulation took
4.89 seconds per frame and a total of 11.33 GB in memory.

\begin{figure}
    \centering
    \includegraphics[width=\linewidth,trim=0 20 0 0]{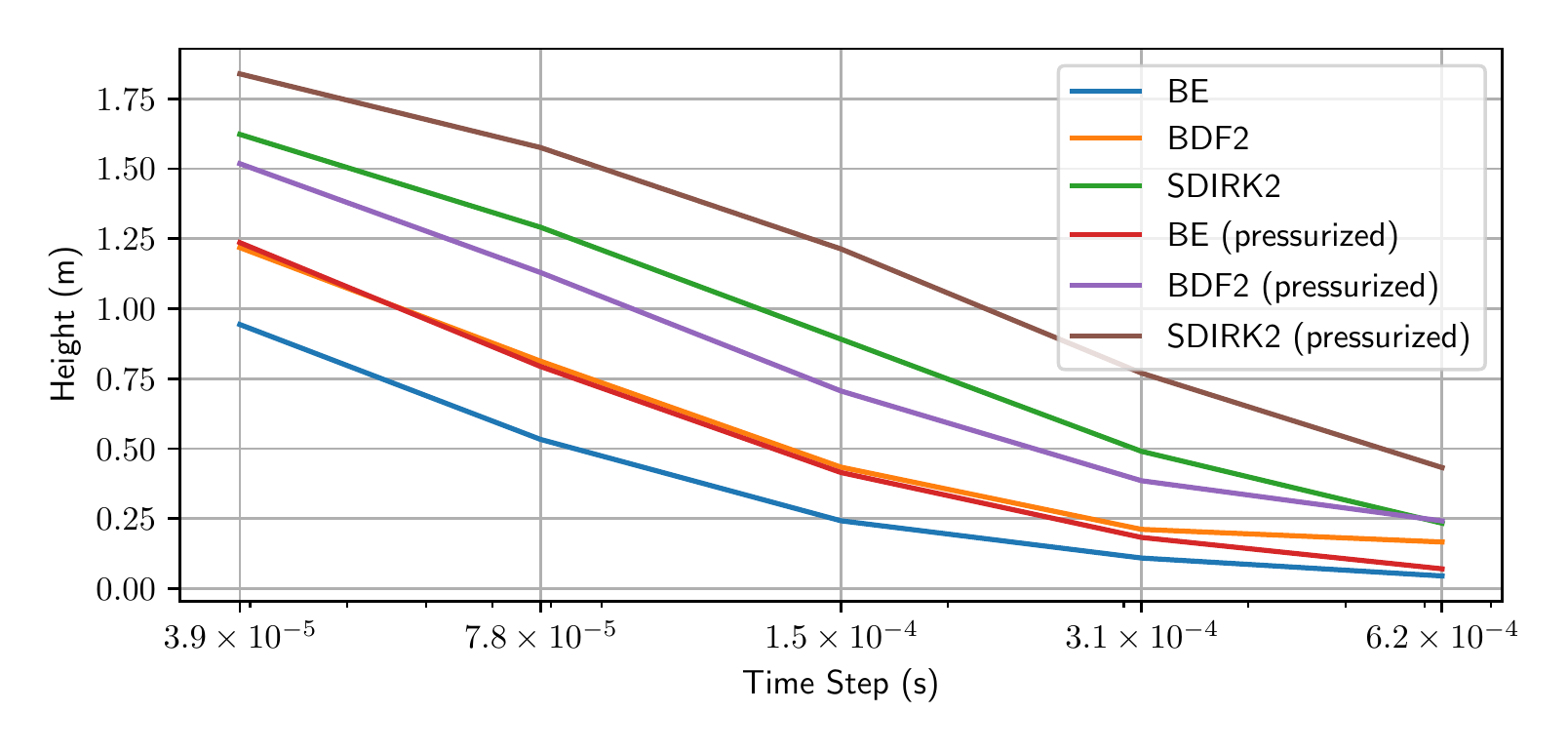}
    \caption{\emph{Tennis ball drop.} A pressurized and non-pressurized tennis
    ball dropped from a 254 cm height, is simulated
    using different integrators. As expected, the pressurized ball bounces higher than
    the corresponding non-pressurized ball. The highly damping BE and BDF2 produce a much lower bounce than
    SDIRK2 across a range of different time steps \cite{ascher21}.
    Higher-order integrators are defined in the supplemental document.}\label{fig:ball_drop}
\end{figure}

Next, a tennis ball is dropped from a 254 cm height to evaluate its bounce with and
without pressurization. We show
how pressurization and choice of integrator can drastically affect the height
of the bounce in \figref{fig:ball_drop}.

\subsubsection{Tire wrinkling}\label{sec:tire_wrinkle}
\begin{figure}[ht]
\centering
    \includegraphics[width=0.5\linewidth,trim=400 100 460 0,clip]{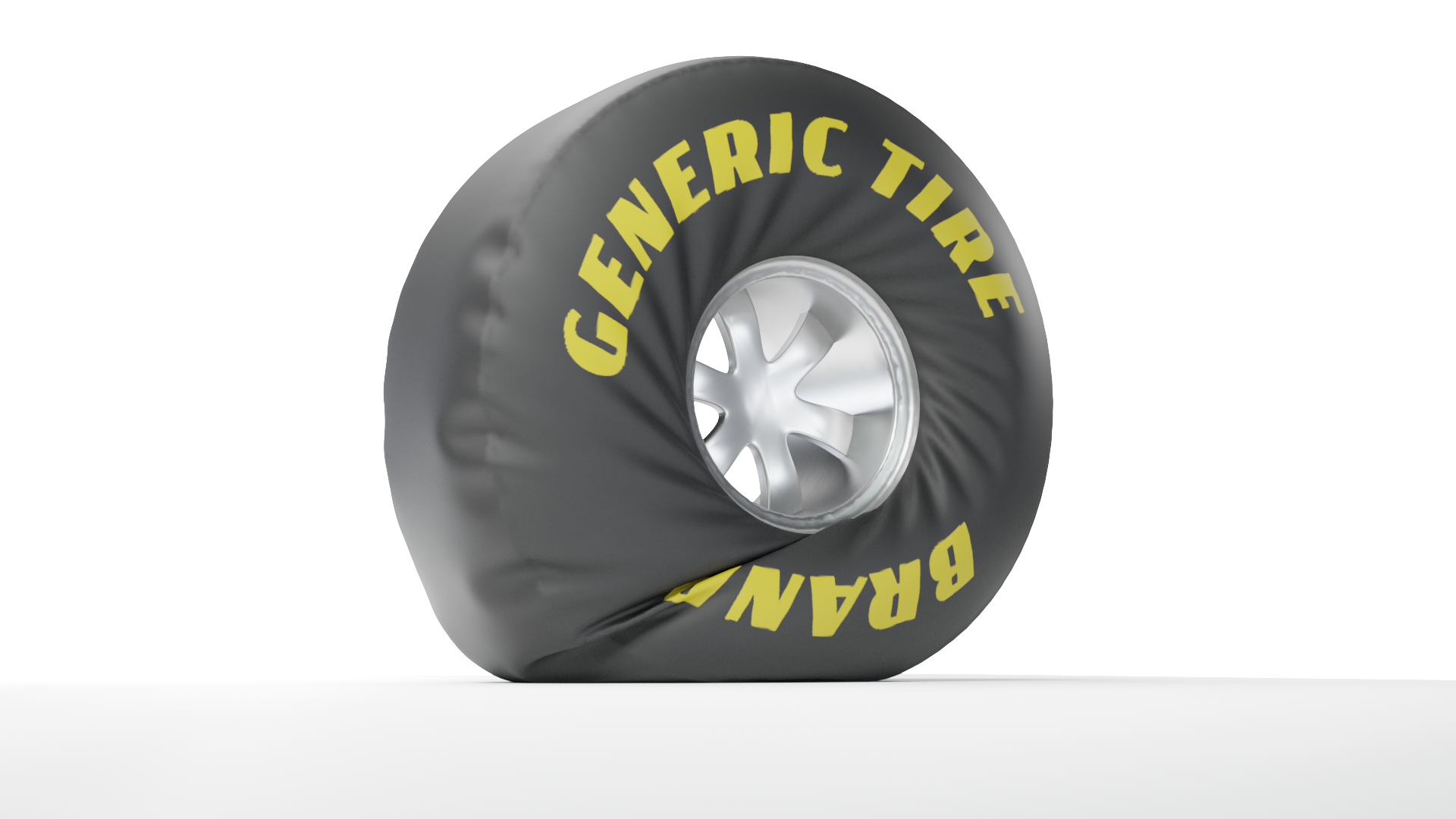}%
    \caption{\emph{Dragster tire wrinkle.}
    A simulation of a soft tire as it spins against the ground with a large friction coefficient,
    causing the rubber to wrinkle.
    The tire is being rotated in place causing the wrinkle
    due to a high coefficient of friction with the ground.
    Real dragster tires exhibit this wrinkle phenomenon at the
    start of the race \cite{goodyear20}.
    \label{fig:tire_folding}}
\end{figure}

We model an inflated tire used by top fuel dragsters to show the folding
phenomenon at the start of the race.  The tires are deliberately inflated at a low pressure of
0.68 atm above atmospheric pressure, which allows them to better grip the asphalt for a better
head start.  As a result the soft tire tends to wrinkle as the wheels start to turn.
This phenomenon allows for a larger contact patch between the tire and the ground for better traction, which
translates to a larger acceleration.
In~\figref{fig:tire_folding} we demonstrate this phenomenon in simulation with a
shell model tire inflated using our volume change penalty.
The outer side of the tire is initially stuck to the ground and then dragged while
maintaining consistent contact.  Accurate simulation of stick-slip transitions of the tire
is critical in determining its performance since traction transfers torque into
forward acceleration of the vehicle, which ultimately determines the outcome of a race.
The tire is simulated using 93K triangles and 46K vertices. The tire mesh is
split into an outer stiffer part that is in contact with the ground and a softer inner part
where the label is printed. The inner rim sets the Dirichlet boundary condition generating the rotation.
Here \new{$\mu_d = 0.5$, $\mu_s = 1.5$}, and $\nu = 0.49$ everywhere, while $E= 400$ KN/m, $\rho = 200$ kg/m${}^2$
and bending stiffness set at \new{$0.56$} on the outer part, and $E = 200$ KN/m, $\rho = 50$ kg/m${}^2$, and bending stiffness
set at $0.01$ on the inner part.
The simulation ran with $h = 0.00125$ s for $20$ seconds per frame on average using the damped Newton solver.
\new{This example shows how the stick-slip phenomenon can be modeled with distinct
static and dynamic friction coefficients using the smoothed Stribeck model proposed in \secref{sec:friction}.}

\section{Conclusions and Limitations} \label{sec:conclusion}

We have presented a fully implicit method for simulating hyperelastic objects subject
to frictional contacts.  This method generalizes the popular optimization
framework for simulating hyperelastics with contact and lagged friction
potentials.  We demonstrate how higher-order integrators can be applied
in our method as well as in IPC-style frameworks.
Our method addresses the lack of friction convergence in lagged friction
formulations by evaluating contacts, friction forces as well as tangential bases
implicitly.

\new{We have extended this approach to include static friction and physically-based
volume preservation that can be used to simulate compressible as well as
nearly incompressible media in a single solve.}

\paragraph{Local minima}
Solving for roots of nonlinear momentum equations allows one to resolve friction
forces accurately, however, this comes with a trade-off.  Optimization theory 
allows one to reliably find a descent direction even when the
objective Hessian is indefinite via projection or filtering techniques.
Although computing the descent direction for finding roots of nonlinear equations
allows one to use the entire unfiltered Jacobian, global convergence can only be
theoretically guaranteed when the Jacobian is bounded on the neighborhood of the
initial point.  For stiff systems, this assumption can become problematic,
although the practical implications are unclear.

\paragraph{Hydrostatic equilibrium}
Our volume change penalty model expects hydrostatic equilibrium, which may not always be
a good approximation.  For quickly
deforming objects like in the tennis ball and tire examples, some details of
the deformation may be missing due to this approximation. This is because the object deforms
faster than the air moves inside the volume, creating a non-uniform pressure distribution.
The comparison of our hydrostatic model to a fully dynamic fluid simulation remains
as future work.

% Bibliography
\bibliographystyle{IEEEtran}
\bibliography{tvcg}

\end{document}